\documentclass[prl,amsmath,amssymb, notitlepage, twocolumn,floatfix,nofootinbib,superscriptaddress, longbibliography]{revtex4-2}
\usepackage{amsmath}
\usepackage{mathtools} 
\usepackage{tabularx,graphicx}
\usepackage{epstopdf}
\usepackage{txfonts}
\usepackage{graphicx}
\usepackage{enumerate}
\usepackage{spectralsequences} 
\usepackage{latexsym}
\usepackage{amssymb}
\usepackage{tikz-cd}
\usepackage{amsmath}
\usepackage{color, colortbl}
\usepackage{psfrag}
\usepackage{bbm}
\usepackage{bm}
\usepackage{titlesec}
\usepackage{feynmp}

\usepackage{slashed}
\usepackage{multirow}
\usepackage{framed}
\usepackage{braket}
\usepackage{comment}
\usepackage{appendix}

\newcommand{\beq}{\begin{eqnarray}}
	\newcommand{\eeq}{\end{eqnarray}}

\newcommand{\Z}{\mathbb{Z}}

\newcommand{\bs}{\boldsymbol}

\DeclareMathOperator{\tr}{tr}

\newcommand{\bsp}{\begin{split}}
	\newcommand{\esp}{\end{split}}

\newcommand{\E}{\mathcal{E}}

\usepackage{color}
\definecolor{darkblue}{rgb}{0.,0.,0.4}
\definecolor{darkred}{rgb}{0.5,0.,0.}
\definecolor{BlueViolet}{RGB}{138,43,226}
\definecolor{SkyBlue}{RGB}{30,144,255}
\definecolor{DarkGreen}{RGB}{0,100,0}
\usepackage[pdftex,colorlinks=true,linkcolor=violet,citecolor=blue,urlcolor=darkred]{hyperref}
\usepackage[normalem]{ulem}

\makeatletter
\newsavebox{\@brx}
\newcommand{\llangle}[1][]{\savebox{\@brx}{\(\m@th{#1\langle}\)}%
  \mathopen{\copy\@brx\kern-0.5\wd\@brx\usebox{\@brx}}}
\newcommand{\rrangle}[1][]{\savebox{\@brx}{\(\m@th{#1\rangle}\)}%
  \mathclose{\copy\@brx\kern-0.5\wd\@brx\usebox{\@brx}}}
\makeatother
\usepackage{amsthm}
\theoremstyle{plain}
\newtheorem*{theorem*}{Theorem}

\makeatletter
\newcommand*{\rom}[1]{\expandafter\@slowromancap\romannumeral #1@}
\makeatother

\renewcommand{\vec}[1]{\bm{#1}}

\begin{document}
\title{Fluctuation-Dissipation Theorem and Information Geometry in Open Quantum Systems}
	
\author{Jian-Hao Zhang}
\affiliation{Department of Physics and Center for Theory of Quantum Matter, University of Colorado, Boulder, Colorado 80309, USA}

\author{Cenke Xu}
\email{xucenke@physics.ucsb.edu}
\affiliation{Department of Physics, University of California, Santa Barbara, CA 93106, USA}

\author{Yichen Xu}
\email{yx639@cornell.edu}
\affiliation{Department of Physics, Cornell University, Ithaca, New York 14850, USA}

\begin{abstract}

We propose a fluctuation-dissipation theorem in open quantum systems from an information-theoretic perspective. We define the fidelity susceptibility that measures the sensitivity of the systems under perturbation and relate it to the fidelity correlator that characterizes the correlation behaviors for mixed quantum states. In particular, we determine the scaling behavior of the fidelity susceptibility in the strong-to-weak spontaneous symmetry breaking (SW-SSB) phase, strongly symmetric short-range correlated phase, and the quantum critical point between them.  We then provide a geometric perspective of our construction using distance measures of density matrices. We find that the metric of the quantum information geometry generated by perturbative distance between density matrices before and after perturbation is generally non-analytic. Finally, we design a polynomial proxy that can in principle be used as an experimental probe for detecting the SW-SSB and phase transition through quantum metrology. In particular, we show that each term of the polynomial proxy is related to the R\'enyi versions of the fidelity correlators.

\end{abstract}

\maketitle

\textit{Introduction} -- The concept of \textit{spontaneous symmetry breaking} (SSB) is one of the organizing principles in modern condensed matter physics \cite{LandauLifshitz, McGreevy2022}. SSB in pure state and thermal states are well-understood. Meanwhile, a mixed quantum state $\rho$ admits two classes of global symmetries \cite{deGroot2022, MaWangASPT, ZhangQiBi2022, LeeYouXu2022, Albert2018, AlbertJiang2014, Lee_2023, chen2023symmetryenforced, chen2024separability, chen2024unconventional, Kawabata_2024, zhou2023reviving, Zhang_2023, ma2024symmetry, xue2024tensor, guo2024locally, chirame2024stable, xu2024average, hsin2023anomalies, lessa2024mixedstate, wang2024anomaly, guo2024designing, guo2024new}: one is the \textit{strong symmetry} defined as $U\rho=e^{i\theta}\rho$ where $U$ is the global symmetry transformation, i.e. every eigenstate of the density matrix carries the same charge under the symmetry; the other is the \textit{weak symmetry}, defined as $U\rho U^\dag=\rho$, which means that different eigenstates of the density matrix are still the eigenstates of the global symmetry charge, but they may carry different eigenvalues of the symmetry charge. Recently, a new type of SSB pattern, dubbed \textit{strong-to-weak SSB} (SW-SSB), has been found in mixed states \cite{Lee_2023, ma2024topological, lessa2024strong, sala2024spontaneous, xu2024average, huang2024hydro, gu2024spontaneous, moharramipour2024symmetry, kuno2024strong, su2024emergent, sala2024decoherence, zhang2024strong}, where a strong global symmetry is spontaneously broken down to a weak one. 
It was proposed that the most proper definition of the SW-SSB state~\cite{lessa2024strong} is through the fidelity correlator, namely\footnote{Ref. \cite{lessa2024strong} used the square root fidelity correlator to define the SWSSB phase, which is slightly different but equivalent to the present definition.}
\begin{align}
F\left(\rho,O_iO_j^\dag\rho O_jO_i^\dag\right)\sim O(1),
\label{Eq: fidelity correlator}
\end{align}
where $F(\rho, \sigma)=(\tr\sqrt{\sqrt{\rho}\sigma\sqrt{\rho}})^2$ is the fidelity of two mixed-state density matrices $\rho$ and $\sigma$ and $O_i$ is a local operator carrying charges of the strong symmetry. This definition through fidelity correlator ensures that the SW-SSB state is a robust mixed state phase, in the sense that it cannot be ``two-way" connected to a symmetric state through symmetric finite or log-depth local quantum channels \cite{lessa2024strong}. Meanwhile, the linear correlator $\tr\left(\rho O_i^\dagger O_j\right)$, broadly used as the diagnosis for SSB in pure and mixed states, is short-ranged for the SW-SSB states. 

In quantum many-body physics, symmetry-breaking orders are rigorously studied via the linear response theory \cite{Sachdev_2011,mahan2013many}, where the sensitivity of the order parameter to external perturbation, i.e. the susceptibility, is proportional to the linear correlator of local charged operators. This is colloquially known as the fluctuation-dissipation theorem \cite{callen1951irreversibility,kubo1966fluctuation,marconi2008fluctuation}. 
However, as the linear correlator for the SW-SSB phase is short-ranged, such an ordered state will not be detected through the conventional linear response theory. This calls for a new formalism to probe mixed quantum states with strong symmetries.

We begin by defining a fidelity-based ``magnetization",  whose susceptibility under external perturbation via a quantum channel, is then shown to be related to the fidelity correlator \eqref{Eq: fidelity correlator}\footnote{We note that the term ``fidelity susceptibility" has already been widely used in literature \cite{cozzini2007quantum,you2007fidelity,chen2008intrinsic,zhou2008fidelity,quan2009quantum,gu2010fidelity,albuquerque2010quantum,banchi2014quantum,carollo2018uhlmann,carollo2020geometry}. The major difference is that the fidelity susceptibility in our case is the sensitivity to perturbation via external quantum channel, rather than perturbation in Hamiltonian. }. This generalizes the fluctuation-dissipation theorem on the level of fidelity. The fidelity susceptibility diverges linearly with increasing system size in the SW-SSB phase and remains finite in the symmetric short-range correlated phase. 
At the quantum critical point or intermediate phase between the two phases, the fidelity susceptibility either remains finite or scales sub-linearly with system size.

We then consider the response of the mixed state to external perturbation from the viewpoint of quantum information geometry, where the effect of the perturbation is quantified by the Bures distance between perturbed and unperturbed density matrices. We find that the metric of the information geometry is generally non-analytic, and encodes the correlation behavior of the state. 
Finally, we provide proxies of the fidelity susceptibility that involve only polynomials of the density matrix $\rho$, which facilitates experimental measurement of the fidelity susceptibility.

\textit{Fidelity magnetization and susceptibility} -- Motivated by the form of the fidelity correlator \eqref{Eq: fidelity correlator}, we define the fidelity magnetization of a mixed-state $\rho$ as
\begin{align}
M_F(\rho):=\frac{1}{N}\sum_iF\left(\rho, O_i^\dag \rho O_i\right),
\label{Eq: fidelity magnetization}
\end{align}
where $N$ is the total number of sites in the system, and $O^\dagger_i$ is the charge creation operator. In this work, we will take the U(1) or $\Z_n$ symmetry as examples, as the charge creation operator $O^\dagger$ of the U(1) and $\Z_n$ symmetry can also serve as the order parameter of the symmetry, and they can be made {\it unitary}, which ensures that $O^\dagger_i \rho O_i$ is a valid density matrix. If $\rho=\ket{\psi}\bra{\psi}$ is a pure state, the fidelity magnetization reduces to the square of traditional magnetization: $\frac{1}{N}\sum_i|\bra{\psi}O_i^\dag\ket{\psi}|^2$. We subsequently define the fidelity susceptibility as the change of the fidelity magnetization when $\rho$ is perturbed by the following local charge-dephasing quantum channel: 
\begin{align}
\E=\circ_j\E_j,~\E_j[\rho]=(1-p)\rho+p O_j \rho O_j^\dag,
\label{Eq: channel}
\end{align}
which reads
\begin{align}
\label{Eq: fidelity susceptibility}
\chi_F&=\lim_{p\rightarrow 0}\frac{M_F(\mathcal{E}(\rho))-M_F(\rho)}{p}\nonumber\\
&=\frac{1}{N}\sum_i\lim_{p\rightarrow 0}\frac{F\left(\E[\rho], O_i^\dag \E[\rho] O_i\right)-F\left(\rho, O_i^\dag \rho O_i\right)}{p}.
\end{align}
The unitarity of $O_i$ ensures that $O^\dagger_i \rho O_i$ is a valid density matrix, and Eq.~\eqref{Eq: channel} is a valid quantum channel.

For a density matrix with strong global symmetry, the fidelity magnetization $M_F(\rho)$ vanishes, and the fidelity susceptibility can be computed as
\begin{widetext}
\begin{align}
\label{Eq: perturbed fidelity susceptibility}
\chi_F&=\lim_{p\to 0}\frac{1}{Np}\sum_i \left[F\left((1-p)^N\rho+p(1-p)^{N-1}\sum\limits_jO_j\rho O_j^\dag, \ \ (1-p)^{N} O_i^\dag\rho O_i+p(1-p)^{N-1}\sum_{j}O_i^\dag O_j\rho O_j^\dag O_i\right)+O(p^2)\right]\\
&= \lim_{p\to 0}(1-p)^{2N-1} \sum_i\left[\sqrt{F}\left(\rho,\frac{1}{N}\sum_{j}O_i^\dag O_j\rho O_j^\dag O_i\right)+\sqrt{F}\left(\frac{1}{N}\sum_jO_j\rho O_j^\dag, O_i^\dag \rho O_i\right)\right]^2=\eta\sum_iF\left(\rho,\frac{1}{N}\sum_{j}O_i^\dag O_j\rho O_j^\dag O_i\right),\nonumber
\end{align}
\end{widetext}
where we use the additivity of fidelity between density matrices with strong symmetries, and take square of Eq.S1 in the supplementary material (SM)~\cite{supplementary}. We note here that we have assumed $p \ll 1/N$ in the derivation. The factor $\eta=4$ 
when $O$ is a $\Z_2$ order parameter (i.e., $O^\dag=O$), for instance when $O$ is the charge-$n$ creation operator of a $Z_{2n}$ symmetry, otherwise $\eta=1$; for more details please refer to the SM~\cite{supplementary}.

Eq. \eqref{Eq: perturbed fidelity susceptibility} draws a clear analogy with the fluctuation-dissipation theorem that holds in a variety of physical scenarios: the susceptibility of the fidelity magnetization is related to the fidelity correlator of the operator $O$ across different positions, which quantifies the fluctuation of the mixed state $\rho$ with respect to external channel perturbation. We also note that another definition of susceptibility was given based on the classical Fisher information of a density matrix $\rho$ with respect to an infinitesimal decoherence channel analogous to Eq.~\eqref{Eq: channel}, and the connection to the R\'enyi correlator and correlation in the doubled space was made~\cite{Lee_2023}.

\textit{Fidelity susceptibility of mixed-state phases} -- We now examine the behavior of the fidelity susceptibility in various mixed-state quantum phases. To proceed, we point out that the fidelity susceptibility can be both lower and upper bounded by the sum of the (square-root) fidelity correlator over the entire system (from now on, we will always assume that the system is translation invariant, therefore the fidelity in the summation in Eq. \eqref{Eq: perturbed fidelity susceptibility} is independent of $i$ ):
\begin{align}\label{eq: chifinequal}
\sum_jF\left(\rho, O_i^\dag O_j \rho O_j^\dag O_i\right)\leq \chi_F\leq \left[\sum_j\sqrt{F}\left(\rho, O_i^\dag O_j \rho O_j^\dag O_i\right)\right]^2,
\end{align}
where the lower bound is guaranteed by the concavity of fidelity, and the upper bound is proved in the SM~\cite{supplementary}. 

We first discuss the behavior of the fidelity susceptibility in the SSB phase of a global symmetry, including the SW-SSB phase, an intrinsically mixed-state phase with no pure-state analogue. The SSB order is characterized by the long-ranged fidelity correlator \eqref{Eq: fidelity correlator}. In the SSB phase, the lower bound of $\chi_F$ in Eq. \eqref{eq: chifinequal} scales as $\sim O(N)$ due to the long-range nature of the fidelity correlator \eqref{Eq: fidelity correlator}, and $\chi_F$ itself is a sum of fidelity over the whole system, which is upper bounded by $N$. Therefore, the susceptibility diverges in the thermodynamic limit $N\to\infty$ with the scaling behavior $\chi_F\sim O(N)$. Physically, the divergent susceptibility implies that a local channel perturbation would induce a global change in fidelity magnetization, a clear indication of the spontaneity of symmetry breaking.

On the other hand, we show that the susceptibility does not scale with the system size $N$ in a symmetric short-range correlated (SRC) phase (i.e. the paramagnetic phase) where the fidelity correlator decays exponentially, namely
\begin{align}
F\left(\rho,O_i^\dag O_j \rho O_j^\dag O_i\right)\sim e^{-|i-j|/\xi},
\label{Eq: SRC fidelity}
\end{align}
where $|i-j|$ is the spatial distance between $i$ and $j$ and $\xi$ is defined as the \textit{fidelity correlation length}. In fact, the lower and upper bounds in Eq. \eqref{eq: chifinequal} are both finite:
\begin{align}
\chi_F\geq\sum_jF\left(\rho, O_i^\dag O_j \rho O_j^\dag O_i\right)\sim \int\frac{\mathrm{d}^d\bs{x}}{a^d}e^{-|\bs{x}|/\xi}\sim\left(\frac{\xi}{a}\right)^d,
\end{align}
where we take the continuous limit $\sum_j\to \int \frac{\mathrm{d}^d\bs{x}}{a^d}$ and $a$ is a typical lattice constant. Meanwhile,
\begin{align}
\chi_F\leq\left[\sum_j\sqrt{F}\left(\rho, O_i^\dag O_j \rho O_j^\dag O_i\right)\right]^2\sim\left(\int\frac{\mathrm{d}^d\bs{x}}{a^d}e^{-|\bs{x}|/2\xi}\right)^2\sim \left(\frac{2\xi}{a}\right)^{2d}.
\end{align}
Both bounds are finite in the SRC phase, where $\xi/a\sim O(1)$.

In principle, the SW-SSB and SRC phases could be separated by either a critical point or an intermediate phase. Here we assume that, in such situation, the fidelity correlator decays algebraically, namely
\begin{align}
F\left(\rho, O_i^\dag O_j \rho O_j^\dag O_i\right)\sim |i-j|^{-\gamma},
\label{Eq: critical fidelity}
\end{align}
where $\gamma$ is some positive exponent. The lower and upper bounds by Eq. \eqref{eq: chifinequal} now depend on $\gamma$, namely
\begin{align}
\chi_F\geq\sum_jF\left(\rho, O_i^\dag O_j \rho O_j^\dag O_i\right)\sim \int\frac{\mathrm{d}^d\bs{x}}{a^d}|\bs{x}|^{-\gamma}\sim \left\{ \begin{array}{rrr}
 O(1),\ \gamma>d,\\
 \log N,\ \gamma=d,\\
N^{1-\frac{\gamma}{d}},\ \gamma<d,
\end{array}\right.
\end{align}
and
\begin{align}
\chi_F&\leq\left[\sum_j\sqrt{F}\left(\rho, O_i^\dag O_j \rho O_j^\dag O_i\right)\right]^2\nonumber\\
&\sim\left(\int\frac{\mathrm{d}^d\bs{x}}{a^d}|\bs{x}|^{-\gamma/2}\right)^2\sim \left\{ \begin{array}{rrr}
 O(1),\ \gamma>2d,\\
 \log^2 N,\ \gamma=2d,\\
N^{2-\frac{\gamma}{d}},\ \gamma < 2d.
\end{array}\right.
\end{align}
Hence when $\gamma>2d$, $\chi_F\sim O(1)$, and when $\gamma<d$, $\chi_F$ scales sub-linearly with the system size $N$ as
\begin{equation}
    N^{1-\frac{\gamma}{d}}\leq \chi_F\leq N.
\end{equation}
When $d\leq \gamma\leq 2d$, the exact scaling behavior of the fidelity susceptibility in the intermediate phase depends on the physical details.

\textit{Quantum information geometry and susceptibility} -- 
We now discuss an alternative quantification of the mixed state's response to external channel perturbation from the perspective of quantum information geometry, where the perturbation is quantified by geometric distances between perturbed and unperturbed density matrices. A broadly used distance measure between density matrices is the Bures distance, whose square is defined as \cite{bures1969extension,Uhlmann,hubner1992explicit,braunstein1994statistical,paris2009quantum,holevo2011probabilistic,bengtsson2017geometry, 9965836}
\begin{align}
D_b^2(\rho,\sigma)=2-2\sqrt{F}(\rho,\sigma),
\label{Eq: Bures distance}
\end{align}
where $\rho$ and $\sigma$ are two density matrices and $\sqrt{F}$ is the square root fidelity. The second order expansion of the Bures distance $D_b^2(\rho,\rho+d\rho)$ between a density matrix $\rho$ and a perturbed density matrix $\rho+d\rho$ is known as the Bures metric, which describes the susceptibility of the Bures distance under perturbation, and is generally linked to the physical susceptibility when $d\rho$ is generated by an external perturbation. Barring a sudden change of the rank of $\rho$ under perturbation, the quantum fisher information metric, the central quantity in quantum metrology, is four times the Bures metric\cite{vsafranek2017discontinuities,zhou2019exact,meyer2021fisher}.

In our case, we define a general space-varying channel perturbation that generalizes Eq. \eqref{Eq: channel}:
\begin{align}
    \mathcal{E}_{\vec{\theta}}\equiv \circ_j\mathcal{E}_j,\ \mathcal{E}_j[\rho]=\cos^2\theta_j\rho+\sin^2\theta_j O_j\rho O_j^\dagger,
\end{align}
where $\vec{\theta}=(\theta_1,\theta_2,\dots,\theta_N)$ is an $N$-dimensional vector that parameterizes the channel perturbation as infinitesimal rotations. For convenience, we always assume $\theta_i>0$. To characterize the geometry of the mixed states generated by such a parameterized channel, we first consider the following distance $D_b^2\left(\mathcal{E}_{\vec{\theta}}[\rho],\rho\right)$. However, a straightforward calculation shows that, upon expanding to the second order of $\theta$, we always have
\begin{align}
    D_b^2\left(\mathcal{E}_{\vec{\theta}}[\rho],\rho\right)=\sum_i\theta_i^2=|\vec{\theta}|^2,
\end{align}
regardless of physical details of $\rho$. This is simply because $\rho$ has strong symmetry, hence every charged state at the $\theta^2$ order generated by the channel perturbation is always orthogonal to the original one, reminiscent of the ``orthogonality catastrophe" \cite{anderson}. 
Instead, we consider the following quantity between two perturbed density matrices
\begin{equation}\label{eq: dthetaphi}
    g_B\left(\vec{\theta},\vec{\phi}\right)\equiv \frac{D_b^2\left(\mathcal{E}_{\vec{\theta}}[\rho],\rho\right)+D_b^2\left(\rho,\mathcal{E}_{\vec{\phi}}[\rho]\right)-D_b^2\left(\mathcal{E}_{\vec{\theta}}[\rho],\mathcal{E}_{\vec{\phi}}[\rho]\right)}{2}.
\end{equation}
If $\vec{\theta}$ and $\vec{\phi}$ were to parameterize a Riemannian manifold, expanding Eq. \eqref{eq: dthetaphi} to the second order of $\theta$ and $\phi$ would yield the metric that defines the inner product of such a manifold: $g_B\left(\vec{\theta},\vec{\phi}\right)\sim \sum_{ij}g_{ij}\theta_i\phi_j$, where $g_{ij}$ is the Riemannian metric. However, we will show that this is only the case when the system is at the SRC fixed point. In fact, away from the SRC fixed point, the metric will generally be a non-analytic function of $\theta_i$ and $\phi_i$. 

\begin{figure}
    \centering
    \includegraphics[width=0.45\textwidth]{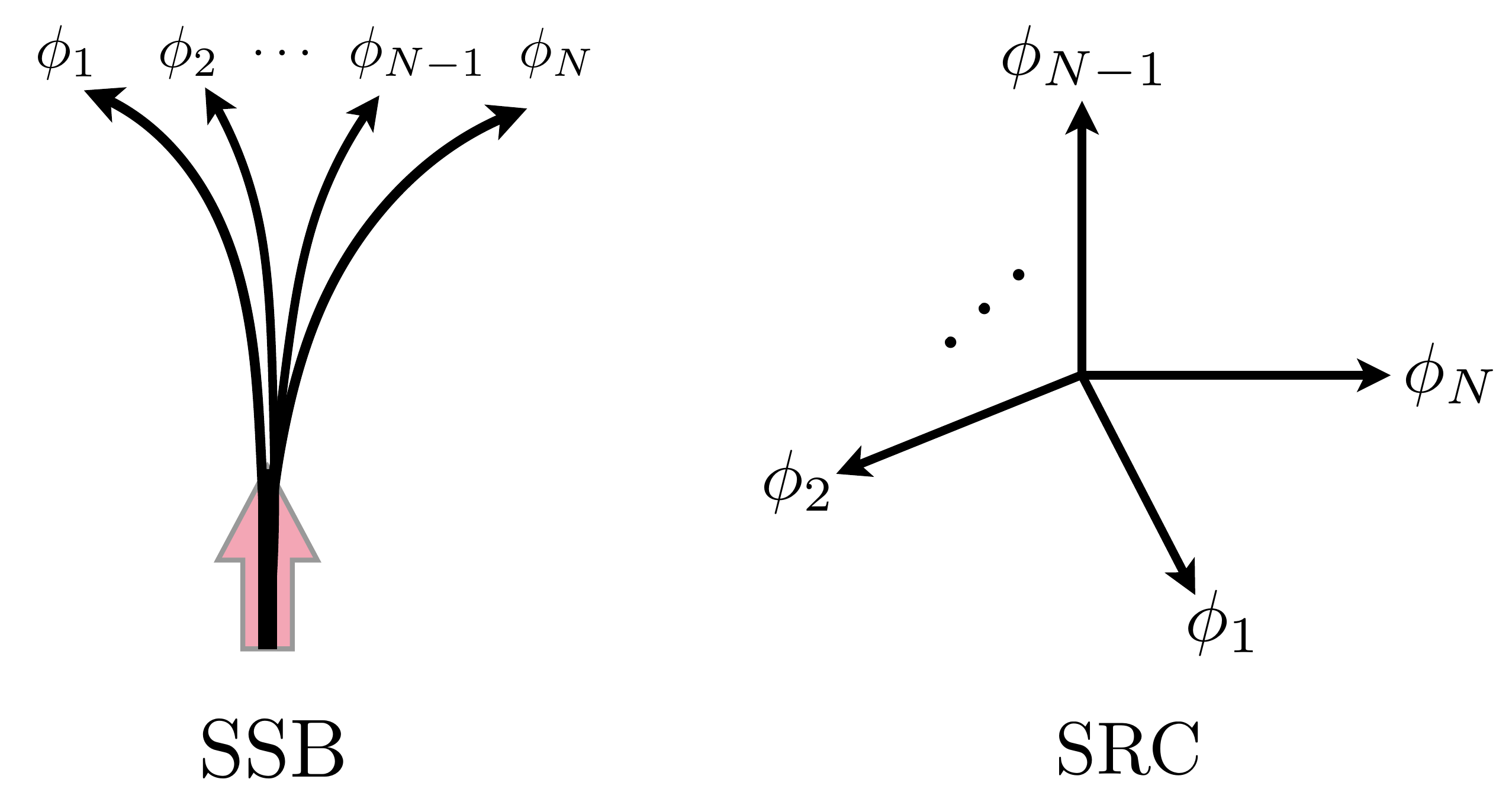}
    \caption{Schematic illustration of quantum information geometry generated by the perturbative channel $\E_{\vec{\phi}}$ in SSB and SRC phases. In the SSB phase, the effective dimension $D_\text{eff}$ of the geometry is almost 1-dimensional, since perturbing in any direction $\phi_i$ yields the same distance [cf. Eq. \eqref{eq: ssbmetric}], and in the SRC phase different perturbative directions are mostly orthogonal to each other [cf. Eq. \eqref{eq: srcmetric}].}
    \label{fig: SSB_SRC}
\end{figure}

We investigate the second-order expansion of the Bures metric
\begin{align}\label{eq: buresmetric}
g_B\left(\vec{\theta},\vec{\phi}\right)=\sqrt{F}\left(\sum_i\theta_i^2O_i\rho O_i^\dagger, \sum_j\phi_j^2 O_j\rho O_j^\dagger\right).
\end{align} To directly visualize the non-analyticity, we consider the following metric
\begin{align}\label{eq: gb2}
g_B\left(\vec{\theta}^i,\vec{\phi}^j+\vec{\phi}^k\right)=\sqrt{F}\left(\theta_i^2O_i\rho O_i^\dagger, \phi_j^2 O_j\rho O_j^\dagger+\phi_k^2 O_k\rho O_k^\dagger\right),
\end{align}
where $(\vec{\theta}^i)_k=\theta_i\delta_{ik}$ and $(\vec{\phi}^j)_k=\phi_j\delta_{jk}$ represent perturbations on one single site. The physical picture of $g_B\left(\vec{\theta}^i,\vec{\phi}^j+\vec{\phi}^k\right)$ is to turn on two independent perturbations at $j$ and $k$ and measure the system's response to them at $i$. Using the convexity of Bures distance, we have
\begin{align}\label{eq: fidelityinnerinequal}    g_B\left(\vec{\theta}^i,\vec{\phi}^j+\vec{\phi}^k\right)\leq g_B\left(\vec{\theta}^i,\vec{\phi}^j\right)+g_B\left(\vec{\theta}^i,\vec{\phi}^k\right).
\end{align}
If $g_B$ in Eq.~\ref{eq: gb2} is an analytic function of $\vec{\theta}$ and $\vec{\phi}$, then by definition it can be expanded as a polynomial of $\theta_i$ and $\phi_i$. However, $g_B$ in Eq.~\ref{eq: gb2} is always linear with the amplitude of $\vec{\theta}$ and $\vec{\phi}$, hence this polynomial can only be a linear function of $\phi_i$, which would make Eq.~\ref{eq: fidelityinnerinequal} an equality rather than inequality. If Eq.~\ref{eq: fidelityinnerinequal} is an inequality, it generally implies that $g_B$ in Eq.~\ref{eq: gb2} cannot be an analytic function of $\vec{\theta}$ and $\vec{\phi}$. The equality of Eq.~\ref{eq: fidelityinnerinequal} holds only when $O_j\rho O_j^\dagger$ is orthogonal to $O_k\rho O_k^\dagger$, i.e. the limit of fixed-point SRC phase. In this case, the Bure metric is
\begin{equation}\label{eq: srcmetric}
g_B\left(\vec{\theta},\vec{\phi}\right)\stackrel{\text{SRC}}{=}\sum_i\theta_i\phi_i=\vec{\theta}\cdot\vec{\phi},
\end{equation}
which is the inner product between two vectors in the $N$-dimensional Euclidean space. Physically, Eq. \eqref{eq: fidelityinnerinequal} shows that the response caused by two independent perturbations is not a simple sum of two individual responses, which would be the case in linear response theory.


The non-analyticity of the Bures metric is most drastic in the fixed-point SSB phase, where $O_i\rho O_i^\dagger=O_j\rho O_j^\dagger$ for every $i$ and $j$. The Bures metric becomes
\begin{equation}\label{eq: ssbmetric}
    g_B\left(\vec{\theta},\vec{\phi}\right)\stackrel{\text{SSB}}{=}|\vec{\theta}||\vec{\phi}|,
\end{equation}
i.e. the entire parameter space collapses into a 1-dimensional flat space. This is because there is only one meaningful perturbation that will be responded by the SSB state: the perturbation that changes the global symmetry charge. We sketch the quantum information geometry of the mixed states generated by the perturbative channel $\E_{\vec{\phi}}$ in Fig. \ref{fig: SSB_SRC}.

In general, we can define the effective dimension of the parameter space as
\begin{equation}
    D_\text{eff}=\frac{N}{\chi_F}.
\end{equation}
Physically, $D_\text{eff}$ counts the number of independent degrees of freedom that are sensitive to the channel perturbation, which can also be roughly estimated as $D_\text{eff}\sim V/\xi^d$, where $V$ is the system volume and $\xi$ is the fidelity correlation length defined in Eq. \eqref{Eq: SRC fidelity}.

\textit{Experimental probe of the fidelity susceptibility -- }The susceptibility defined in Eq. \eqref{Eq: perturbed fidelity susceptibility} involves the fidelity between two density matrices, experimental measurement of the fidelity susceptibility calls for the full quantum state tomography (QST) of the density matrix $\rho$. To practically access the fidelity susceptibility, it is more convenient to measure proxies of the fidelity that are polynomials of the density matrix $\rho$. As shown in Ref. \cite{miszczak2008sub}, the fidelity $F(\rho,\sigma)$ can be lower-bounded by the following inequality:
\begin{align}
F(\rho,\sigma)\geq \tr(\rho\sigma)+\sqrt{2}\sqrt{\left[\tr(\rho\sigma)\right]^2-\tr\left[(\rho\sigma)^2\right]}.
\label{eq: fidelitylower}
\end{align}
Therefore, assuming translation invariance, the fidelity susceptibility can be lower-bounded by
\begin{widetext}
\begin{align}\label{eq: chifproxy}
    \frac{\chi_F}{N}&=F\left(\rho, \frac{1}{N}\sum_j O_i^\dagger O_j\rho O_j^\dagger O_i\right) \geq \frac{1}{N}\sum_j \tr\left(\rho O_i^\dagger O_j\rho O_j^\dagger O_i\right)+\sqrt{2}\sqrt{\left[\frac{1}{N}\sum_j\tr\left(\rho O_i^\dagger O_j\rho O_j^\dagger O_i\right)\right]^2-\tr\left[\left(\frac{1}{N}\sum_j\rho O_i^\dagger O_j\rho O_j^\dagger O_i\right)^2\right]}.
\end{align}
\end{widetext}
Such inequality is useful for experimentally tracking a phase transition from an SRC phase to an SSB phase, as the lower bound only contains quantities that are polynomials of $\rho$, which admit more efficient measurement techniques such as classical shadow tomography \cite{Huang_2020, Elben_2022, hu2022Hamiltonian, hu2023classical}. 

Another important consequence of the lower bound in Eq. \eqref{eq: fidelitylower} is that, the R\'enyi-2 correlator, which is widely used as a theoretical proxy in diagnosing the SRC to SW-SSB phase transition \cite{Lee_2023, ma2024topological, lessa2024strong, huang2024hydro}, tends to overestimate the SRC phase, and underestimate the SW-SSB phase. This is because a long-ranged R\'enyi-2 correlator $\tr\left(\rho O_i^\dagger O_j\rho O_j^\dagger O_i\right)$ is a sufficient condition of having a long-ranged fidelity correlator $F\left(\rho, O_i^\dagger O_j\rho O_j^\dagger O_i\right)$ since the latter is always larger. Here we note that in this paper we focus on the 0-form symmetry\footnote{Ref. \cite{lessa2024strong} constructed a ``counterexample" with vanishing fidelity correlator in the thermodynamic limit but finite normalized R\'enyi-2 correlator $\tr\left(\rho O_i^\dagger O_j\rho O_j^\dagger O_i\right)/\tr\left(\rho^2\right)$. This is not inconsistent with Eq. \eqref{eq: chifproxy}, since the bare R\'enyi-2 correlator vanishes in the thermodynamic limit in this example.}. 


In Supplementary Materials \cite{supplementary}, we show that the lower bound in Eq. \eqref{eq: fidelitylower} can be further improved into an increasing sequence that approximates the fidelity, which allows for more accurate experimental estimation of the fidelity susceptibility. Most crucially, we find that the increasing lower bound involves higher R\'enyi correlators, which are defined as
\begin{equation}
R^{(2n)}\equiv\tr\left[\left(\frac{1}{N}\sum_j\rho O_i^\dagger O_j\rho O_j^\dagger O_i\right)^n\right].
\end{equation}
It is important to point out that the R\'enyi-$2n$ correlator is also an $(n+1)$-point correlator between the operator $O$, hence it captures a higher moment of fluctuation. In contrast, for pure states, the susceptibility is a simple average of two-point correlators over the entire space.

\textit{Conclusion and discussion} -- In this work, we formulate a fluctuation-dissipation theorem in open quantum system, where fluctuations are quantified by information-theoretic measures. In particular, we find that the susceptibility $\chi_F$ of fidelity-based order parameters under external channel perturbation is related to the fidelity correlator, which characterizes the fluctuation in the open quantum system. From the perspective of information geometry, we show that the metric generated by perturbed density matrices encodes correlative behaviors. In particular, it is highly non-analytic when the fidelity susceptibility $\chi_F$ diverges.
Finally, to facilitate experimental detection of SSB and phase transitions in open quantum systems, we propose polynomial proxies of fidelity susceptibility using R\'enyi version of the susceptibility, which can be experimentally accessed via classical shadow from randomized measurements. This avoids exponentially hard QST. 

We end this work with some open questions. It is well-known that the response of a non-equilibrium quantum many-body system can be studied using the Keldysh formalism. Therefore, it would be meaningful to compare our approach to the response of the quantum sector in the Keldysh formalism to external sources. Another interesting direction is to consider the relation between fidelty susceptibility and our ability to recover the channel perturbation $\E$, which is potentially linked to the spontaneity of SSB states \cite{lessa2024strong}. We leave these studies to future works.



\acknowledgments{
\textit{Acknowledgments} -- We thank Yimu Bao, Zhen Bi, Meng Cheng, Eun-Ah Kim, Ethan Lake, Andrew Lucas, Zhu-Xi Luo, Rahul Nandkishore, Shengqi Sang, Shijun Sun, Chong Wang, Zongyuan Wang, and Yi-Zhuang You for the enlightening discussions. JHZ is supported by the U.S. Department of Energy under
Award Number DE-SC0024324. YX acknowledges support from the NSF through OAC-2118310. C.X. is supported by the Simons Foundation through the Simons Investigator program. 
}

\bibliography{Refs}

\providecommand{\noopsort}[1]{}\providecommand{\singleletter}[1]{#1}%
\begin{thebibliography}{71}%
\makeatletter
\providecommand \@ifxundefined [1]{%
 \@ifx{#1\undefined}
}%
\providecommand \@ifnum [1]{%
 \ifnum #1\expandafter \@firstoftwo
 \else \expandafter \@secondoftwo
 \fi
}%
\providecommand \@ifx [1]{%
 \ifx #1\expandafter \@firstoftwo
 \else \expandafter \@secondoftwo
 \fi
}%
\providecommand \natexlab [1]{#1}%
\providecommand \enquote  [1]{``#1''}%
\providecommand \bibnamefont  [1]{#1}%
\providecommand \bibfnamefont [1]{#1}%
\providecommand \citenamefont [1]{#1}%
\providecommand \href@noop [0]{\@secondoftwo}%
\providecommand \href [0]{\begingroup \@sanitize@url \@href}%
\providecommand \@href[1]{\@@startlink{#1}\@@href}%
\providecommand \@@href[1]{\endgroup#1\@@endlink}%
\providecommand \@sanitize@url [0]{\catcode `\\12\catcode `\$12\catcode
  `\&12\catcode `\#12\catcode `\^12\catcode `\_12\catcode `\%12\relax}%
\providecommand \@@startlink[1]{}%
\providecommand \@@endlink[0]{}%
\providecommand \url  [0]{\begingroup\@sanitize@url \@url }%
\providecommand \@url [1]{\endgroup\@href {#1}{\urlprefix }}%
\providecommand \urlprefix  [0]{URL }%
\providecommand \Eprint [0]{\href }%
\providecommand \doibase [0]{https://doi.org/}%
\providecommand \selectlanguage [0]{\@gobble}%
\providecommand \bibinfo  [0]{\@secondoftwo}%
\providecommand \bibfield  [0]{\@secondoftwo}%
\providecommand \translation [1]{[#1]}%
\providecommand \BibitemOpen [0]{}%
\providecommand \bibitemStop [0]{}%
\providecommand \bibitemNoStop [0]{.\EOS\space}%
\providecommand \EOS [0]{\spacefactor3000\relax}%
\providecommand \BibitemShut  [1]{\csname bibitem#1\endcsname}%
\let\auto@bib@innerbib\@empty
\bibitem [{\citenamefont {Landau}\ and\ \citenamefont
  {Lifshitz}(1980)}]{LandauLifshitz}%
  \BibitemOpen
  \bibfield  {author} {\bibinfo {author} {\bibfnamefont {L.~D.}\ \bibnamefont
  {Landau}}\ and\ \bibinfo {author} {\bibfnamefont {E.~M.}\ \bibnamefont
  {Lifshitz}},\ }\href@noop {} {\emph {\bibinfo {title} {{Statistical Physics,
  Part 1}}}},\ \bibinfo {series} {Course of Theoretical Physics}, Vol.~\bibinfo
  {volume} {5}\ (\bibinfo  {publisher} {Butterworth-Heinemann},\ \bibinfo
  {address} {Oxford},\ \bibinfo {year} {1980})\BibitemShut {NoStop}%
\bibitem [{\citenamefont {{McGreevy}}(2022)}]{McGreevy2022}%
  \BibitemOpen
  \bibfield  {author} {\bibinfo {author} {\bibfnamefont {J.}~\bibnamefont
  {{McGreevy}}},\ }\bibfield  {title} {\bibinfo {title} {{Generalized
  Symmetries in Condensed Matter}},\ }\href@noop {} {\bibfield  {journal}
  {\bibinfo  {journal} {arXiv e-prints}\ ,\ \bibinfo {eid} {arXiv:2204.03045}}
  (\bibinfo {year} {2022})},\ \Eprint {https://arxiv.org/abs/2204.03045}
  {arXiv:2204.03045 [cond-mat.str-el]} \BibitemShut {NoStop}%
\bibitem [{\citenamefont {de~Groot}\ \emph {et~al.}(2022)\citenamefont
  {de~Groot}, \citenamefont {Turzillo},\ and\ \citenamefont
  {Schuch}}]{deGroot2022}%
  \BibitemOpen
  \bibfield  {author} {\bibinfo {author} {\bibfnamefont {C.}~\bibnamefont
  {de~Groot}}, \bibinfo {author} {\bibfnamefont {A.}~\bibnamefont {Turzillo}},\
  and\ \bibinfo {author} {\bibfnamefont {N.}~\bibnamefont {Schuch}},\
  }\bibfield  {title} {\bibinfo {title} {{Symmetry Protected Topological Order
  in Open Quantum Systems}},\ }\href
  {https://doi.org/10.22331/q-2022-11-10-856} {\bibfield  {journal} {\bibinfo
  {journal} {Quantum}\ }\textbf {\bibinfo {volume} {6}},\ \bibinfo {pages}
  {856} (\bibinfo {year} {2022})}\BibitemShut {NoStop}%
\bibitem [{\citenamefont {Ma}\ and\ \citenamefont {Wang}(2023)}]{MaWangASPT}%
  \BibitemOpen
  \bibfield  {author} {\bibinfo {author} {\bibfnamefont {R.}~\bibnamefont
  {Ma}}\ and\ \bibinfo {author} {\bibfnamefont {C.}~\bibnamefont {Wang}},\
  }\bibfield  {title} {\bibinfo {title} {Average symmetry-protected topological
  phases},\ }\bibfield  {journal} {\bibinfo  {journal} {Physical Review X}\
  }\textbf {\bibinfo {volume} {13}},\ \href
  {https://doi.org/10.1103/physrevx.13.031016} {10.1103/physrevx.13.031016}
  (\bibinfo {year} {2023})\BibitemShut {NoStop}%
\bibitem [{\citenamefont {{Zhang}}\ \emph {et~al.}(2022)\citenamefont
  {{Zhang}}, \citenamefont {{Qi}},\ and\ \citenamefont {{Bi}}}]{ZhangQiBi2022}%
  \BibitemOpen
  \bibfield  {author} {\bibinfo {author} {\bibfnamefont {J.-H.}\ \bibnamefont
  {{Zhang}}}, \bibinfo {author} {\bibfnamefont {Y.}~\bibnamefont {{Qi}}},\ and\
  \bibinfo {author} {\bibfnamefont {Z.}~\bibnamefont {{Bi}}},\ }\bibfield
  {title} {\bibinfo {title} {{Strange Correlation Function for Average
  Symmetry-Protected Topological Phases}},\ }\href
  {https://doi.org/10.48550/arXiv.2210.17485} {\bibfield  {journal} {\bibinfo
  {journal} {arXiv e-prints}\ ,\ \bibinfo {eid} {arXiv:2210.17485}} (\bibinfo
  {year} {2022})},\ \Eprint {https://arxiv.org/abs/2210.17485}
  {arXiv:2210.17485 [cond-mat.str-el]} \BibitemShut {NoStop}%
\bibitem [{\citenamefont {{Lee}}\ \emph {et~al.}(2022)\citenamefont {{Lee}},
  \citenamefont {{You}},\ and\ \citenamefont {{Xu}}}]{LeeYouXu2022}%
  \BibitemOpen
  \bibfield  {author} {\bibinfo {author} {\bibfnamefont {J.~Y.}\ \bibnamefont
  {{Lee}}}, \bibinfo {author} {\bibfnamefont {Y.-Z.}\ \bibnamefont {{You}}},\
  and\ \bibinfo {author} {\bibfnamefont {C.}~\bibnamefont {{Xu}}},\ }\bibfield
  {title} {\bibinfo {title} {{Symmetry protected topological phases under
  decoherence}},\ }\href {https://doi.org/10.48550/arXiv.2210.16323} {\bibfield
   {journal} {\bibinfo  {journal} {arXiv e-prints}\ ,\ \bibinfo {eid}
  {arXiv:2210.16323}} (\bibinfo {year} {2022})},\ \Eprint
  {https://arxiv.org/abs/2210.16323} {arXiv:2210.16323 [cond-mat.str-el]}
  \BibitemShut {NoStop}%
\bibitem [{\citenamefont {{Albert}}(2018)}]{Albert2018}%
  \BibitemOpen
  \bibfield  {author} {\bibinfo {author} {\bibfnamefont {V.~V.}\ \bibnamefont
  {{Albert}}},\ }\bibfield  {title} {\bibinfo {title} {{Lindbladians with
  multiple steady states: theory and applications}},\ }\href
  {https://doi.org/10.48550/arXiv.1802.00010} {\bibfield  {journal} {\bibinfo
  {journal} {arXiv e-prints}\ ,\ \bibinfo {eid} {arXiv:1802.00010}} (\bibinfo
  {year} {2018})},\ \Eprint {https://arxiv.org/abs/1802.00010}
  {arXiv:1802.00010 [quant-ph]} \BibitemShut {NoStop}%
\bibitem [{\citenamefont {Albert}\ and\ \citenamefont
  {Jiang}(2014)}]{AlbertJiang2014}%
  \BibitemOpen
  \bibfield  {author} {\bibinfo {author} {\bibfnamefont {V.~V.}\ \bibnamefont
  {Albert}}\ and\ \bibinfo {author} {\bibfnamefont {L.}~\bibnamefont {Jiang}},\
  }\bibfield  {title} {\bibinfo {title} {Symmetries and conserved quantities in
  lindblad master equations},\ }\href
  {https://doi.org/10.1103/PhysRevA.89.022118} {\bibfield  {journal} {\bibinfo
  {journal} {Phys. Rev. A}\ }\textbf {\bibinfo {volume} {89}},\ \bibinfo
  {pages} {022118} (\bibinfo {year} {2014})}\BibitemShut {NoStop}%
\bibitem [{\citenamefont {Lee}\ \emph {et~al.}(2023)\citenamefont {Lee},
  \citenamefont {Jian},\ and\ \citenamefont {Xu}}]{Lee_2023}%
  \BibitemOpen
  \bibfield  {author} {\bibinfo {author} {\bibfnamefont {J.~Y.}\ \bibnamefont
  {Lee}}, \bibinfo {author} {\bibfnamefont {C.-M.}\ \bibnamefont {Jian}},\ and\
  \bibinfo {author} {\bibfnamefont {C.}~\bibnamefont {Xu}},\ }\bibfield
  {title} {\bibinfo {title} {Quantum criticality under decoherence or weak
  measurement},\ }\bibfield  {journal} {\bibinfo  {journal} {PRX Quantum}\
  }\textbf {\bibinfo {volume} {4}},\ \href
  {https://doi.org/10.1103/prxquantum.4.030317} {10.1103/prxquantum.4.030317}
  (\bibinfo {year} {2023})\BibitemShut {NoStop}%
\bibitem [{\citenamefont {Chen}\ and\ \citenamefont
  {Grover}(2023)}]{chen2023symmetryenforced}%
  \BibitemOpen
  \bibfield  {author} {\bibinfo {author} {\bibfnamefont {Y.-H.}\ \bibnamefont
  {Chen}}\ and\ \bibinfo {author} {\bibfnamefont {T.}~\bibnamefont {Grover}},\
  }\href@noop {} {\bibinfo {title} {Symmetry-enforced many-body separability
  transitions}} (\bibinfo {year} {2023}),\ \Eprint
  {https://arxiv.org/abs/2310.07286} {arXiv:2310.07286 [quant-ph]} \BibitemShut
  {NoStop}%
\bibitem [{\citenamefont {Chen}\ and\ \citenamefont
  {Grover}(2024{\natexlab{a}})}]{chen2024separability}%
  \BibitemOpen
  \bibfield  {author} {\bibinfo {author} {\bibfnamefont {Y.-H.}\ \bibnamefont
  {Chen}}\ and\ \bibinfo {author} {\bibfnamefont {T.}~\bibnamefont {Grover}},\
  }\href@noop {} {\bibinfo {title} {Separability transitions in topological
  states induced by local decoherence}} (\bibinfo {year}
  {2024}{\natexlab{a}}),\ \Eprint {https://arxiv.org/abs/2309.11879}
  {arXiv:2309.11879 [quant-ph]} \BibitemShut {NoStop}%
\bibitem [{\citenamefont {Chen}\ and\ \citenamefont
  {Grover}(2024{\natexlab{b}})}]{chen2024unconventional}%
  \BibitemOpen
  \bibfield  {author} {\bibinfo {author} {\bibfnamefont {Y.-H.}\ \bibnamefont
  {Chen}}\ and\ \bibinfo {author} {\bibfnamefont {T.}~\bibnamefont {Grover}},\
  }\href@noop {} {\bibinfo {title} {Unconventional topological mixed-state
  transition and critical phase induced by self-dual coherent errors}}
  (\bibinfo {year} {2024}{\natexlab{b}}),\ \Eprint
  {https://arxiv.org/abs/2403.06553} {arXiv:2403.06553 [quant-ph]} \BibitemShut
  {NoStop}%
\bibitem [{\citenamefont {Kawabata}\ \emph {et~al.}(2024)\citenamefont
  {Kawabata}, \citenamefont {Sohal},\ and\ \citenamefont
  {Ryu}}]{Kawabata_2024}%
  \BibitemOpen
  \bibfield  {author} {\bibinfo {author} {\bibfnamefont {K.}~\bibnamefont
  {Kawabata}}, \bibinfo {author} {\bibfnamefont {R.}~\bibnamefont {Sohal}},\
  and\ \bibinfo {author} {\bibfnamefont {S.}~\bibnamefont {Ryu}},\ }\bibfield
  {title} {\bibinfo {title} {Lieb-schultz-mattis theorem in open quantum
  systems},\ }\bibfield  {journal} {\bibinfo  {journal} {Physical Review
  Letters}\ }\textbf {\bibinfo {volume} {132}},\ \href
  {https://doi.org/10.1103/physrevlett.132.070402}
  {10.1103/physrevlett.132.070402} (\bibinfo {year} {2024})\BibitemShut
  {NoStop}%
\bibitem [{\citenamefont {Zhou}\ \emph {et~al.}(2023)\citenamefont {Zhou},
  \citenamefont {Li}, \citenamefont {Zhai}, \citenamefont {Li},\ and\
  \citenamefont {Gu}}]{zhou2023reviving}%
  \BibitemOpen
  \bibfield  {author} {\bibinfo {author} {\bibfnamefont {Y.-N.}\ \bibnamefont
  {Zhou}}, \bibinfo {author} {\bibfnamefont {X.}~\bibnamefont {Li}}, \bibinfo
  {author} {\bibfnamefont {H.}~\bibnamefont {Zhai}}, \bibinfo {author}
  {\bibfnamefont {C.}~\bibnamefont {Li}},\ and\ \bibinfo {author}
  {\bibfnamefont {Y.}~\bibnamefont {Gu}},\ }\href@noop {} {\bibinfo {title}
  {Reviving the lieb-schultz-mattis theorem in open quantum systems}} (\bibinfo
  {year} {2023}),\ \Eprint {https://arxiv.org/abs/2310.01475} {arXiv:2310.01475
  [cond-mat.str-el]} \BibitemShut {NoStop}%
\bibitem [{\citenamefont {Zhang}\ \emph {et~al.}(2023)\citenamefont {Zhang},
  \citenamefont {Ding}, \citenamefont {Yang},\ and\ \citenamefont
  {Bi}}]{Zhang_2023}%
  \BibitemOpen
  \bibfield  {author} {\bibinfo {author} {\bibfnamefont {J.-H.}\ \bibnamefont
  {Zhang}}, \bibinfo {author} {\bibfnamefont {K.}~\bibnamefont {Ding}},
  \bibinfo {author} {\bibfnamefont {S.}~\bibnamefont {Yang}},\ and\ \bibinfo
  {author} {\bibfnamefont {Z.}~\bibnamefont {Bi}},\ }\bibfield  {title}
  {\bibinfo {title} {Fractonic higher-order topological phases in open quantum
  systems},\ }\href {https://doi.org/10.1103/PhysRevB.108.155123} {\bibfield
  {journal} {\bibinfo  {journal} {Phys. Rev. B}\ }\textbf {\bibinfo {volume}
  {108}},\ \bibinfo {pages} {155123} (\bibinfo {year} {2023})}\BibitemShut
  {NoStop}%
\bibitem [{\citenamefont {Ma}\ and\ \citenamefont
  {Turzillo}(2024)}]{ma2024symmetry}%
  \BibitemOpen
  \bibfield  {author} {\bibinfo {author} {\bibfnamefont {R.}~\bibnamefont
  {Ma}}\ and\ \bibinfo {author} {\bibfnamefont {A.}~\bibnamefont {Turzillo}},\
  }\href {https://arxiv.org/abs/2403.13280} {\bibinfo {title} {Symmetry
  protected topological phases of mixed states in the doubled space}} (\bibinfo
  {year} {2024}),\ \Eprint {https://arxiv.org/abs/2403.13280} {arXiv:2403.13280
  [quant-ph]} \BibitemShut {NoStop}%
\bibitem [{\citenamefont {Xue}\ \emph {et~al.}(2024)\citenamefont {Xue},
  \citenamefont {Lee},\ and\ \citenamefont {Bao}}]{xue2024tensor}%
  \BibitemOpen
  \bibfield  {author} {\bibinfo {author} {\bibfnamefont {H.}~\bibnamefont
  {Xue}}, \bibinfo {author} {\bibfnamefont {J.~Y.}\ \bibnamefont {Lee}},\ and\
  \bibinfo {author} {\bibfnamefont {Y.}~\bibnamefont {Bao}},\ }\href
  {https://arxiv.org/abs/2403.17069} {\bibinfo {title} {Tensor network
  formulation of symmetry protected topological phases in mixed states}}
  (\bibinfo {year} {2024}),\ \Eprint {https://arxiv.org/abs/2403.17069}
  {arXiv:2403.17069 [cond-mat.str-el]} \BibitemShut {NoStop}%
\bibitem [{\citenamefont {Guo}\ \emph {et~al.}(2024{\natexlab{a}})\citenamefont
  {Guo}, \citenamefont {Zhang}, \citenamefont {Zhang}, \citenamefont {Yang},\
  and\ \citenamefont {Bi}}]{guo2024locally}%
  \BibitemOpen
  \bibfield  {author} {\bibinfo {author} {\bibfnamefont {Y.}~\bibnamefont
  {Guo}}, \bibinfo {author} {\bibfnamefont {J.-H.}\ \bibnamefont {Zhang}},
  \bibinfo {author} {\bibfnamefont {H.-R.}\ \bibnamefont {Zhang}}, \bibinfo
  {author} {\bibfnamefont {S.}~\bibnamefont {Yang}},\ and\ \bibinfo {author}
  {\bibfnamefont {Z.}~\bibnamefont {Bi}},\ }\href@noop {} {\bibinfo {title}
  {Locally purified density operators for symmetry-protected topological phases
  in mixed states}} (\bibinfo {year} {2024}{\natexlab{a}}),\ \Eprint
  {https://arxiv.org/abs/2403.16978} {arXiv:2403.16978 [cond-mat.str-el]}
  \BibitemShut {NoStop}%
\bibitem [{\citenamefont {Chirame}\ \emph {et~al.}(2024)\citenamefont
  {Chirame}, \citenamefont {Burnell}, \citenamefont {Gopalakrishnan},\ and\
  \citenamefont {Prem}}]{chirame2024stable}%
  \BibitemOpen
  \bibfield  {author} {\bibinfo {author} {\bibfnamefont {S.}~\bibnamefont
  {Chirame}}, \bibinfo {author} {\bibfnamefont {F.~J.}\ \bibnamefont
  {Burnell}}, \bibinfo {author} {\bibfnamefont {S.}~\bibnamefont
  {Gopalakrishnan}},\ and\ \bibinfo {author} {\bibfnamefont {A.}~\bibnamefont
  {Prem}},\ }\href@noop {} {\bibinfo {title} {Stable symmetry-protected
  topological phases in systems with heralded noise}} (\bibinfo {year}
  {2024}),\ \Eprint {https://arxiv.org/abs/2404.16962} {arXiv:2404.16962
  [quant-ph]} \BibitemShut {NoStop}%
\bibitem [{\citenamefont {Xu}\ and\ \citenamefont
  {Jian}(2024)}]{xu2024average}%
  \BibitemOpen
  \bibfield  {author} {\bibinfo {author} {\bibfnamefont {Y.}~\bibnamefont
  {Xu}}\ and\ \bibinfo {author} {\bibfnamefont {C.-M.}\ \bibnamefont {Jian}},\
  }\href {https://arxiv.org/abs/2406.07417} {\bibinfo {title} {Average-exact
  mixed anomalies and compatible phases}} (\bibinfo {year} {2024}),\ \Eprint
  {https://arxiv.org/abs/2406.07417} {arXiv:2406.07417 [cond-mat.str-el]}
  \BibitemShut {NoStop}%
\bibitem [{\citenamefont {Hsin}\ \emph {et~al.}(2023)\citenamefont {Hsin},
  \citenamefont {Luo},\ and\ \citenamefont {Sun}}]{hsin2023anomalies}%
  \BibitemOpen
  \bibfield  {author} {\bibinfo {author} {\bibfnamefont {P.-S.}\ \bibnamefont
  {Hsin}}, \bibinfo {author} {\bibfnamefont {Z.-X.}\ \bibnamefont {Luo}},\ and\
  \bibinfo {author} {\bibfnamefont {H.-Y.}\ \bibnamefont {Sun}},\ }\href@noop
  {} {\bibinfo {title} {Anomalies of average symmetries: Entanglement and open
  quantum systems}} (\bibinfo {year} {2023}),\ \Eprint
  {https://arxiv.org/abs/2312.09074} {arXiv:2312.09074 [cond-mat.str-el]}
  \BibitemShut {NoStop}%
\bibitem [{\citenamefont {Lessa}\ \emph
  {et~al.}(2024{\natexlab{a}})\citenamefont {Lessa}, \citenamefont {Cheng},\
  and\ \citenamefont {Wang}}]{lessa2024mixedstate}%
  \BibitemOpen
  \bibfield  {author} {\bibinfo {author} {\bibfnamefont {L.~A.}\ \bibnamefont
  {Lessa}}, \bibinfo {author} {\bibfnamefont {M.}~\bibnamefont {Cheng}},\ and\
  \bibinfo {author} {\bibfnamefont {C.}~\bibnamefont {Wang}},\ }\href@noop {}
  {\bibinfo {title} {Mixed-state quantum anomaly and multipartite
  entanglement}} (\bibinfo {year} {2024}{\natexlab{a}}),\ \Eprint
  {https://arxiv.org/abs/2401.17357} {arXiv:2401.17357 [cond-mat.str-el]}
  \BibitemShut {NoStop}%
\bibitem [{\citenamefont {Wang}\ and\ \citenamefont
  {Li}(2024)}]{wang2024anomaly}%
  \BibitemOpen
  \bibfield  {author} {\bibinfo {author} {\bibfnamefont {Z.}~\bibnamefont
  {Wang}}\ and\ \bibinfo {author} {\bibfnamefont {L.}~\bibnamefont {Li}},\
  }\href@noop {} {\bibinfo {title} {Anomaly in open quantum systems and its
  implications on mixed-state quantum phases}} (\bibinfo {year} {2024}),\
  \Eprint {https://arxiv.org/abs/2403.14533} {arXiv:2403.14533 [quant-ph]}
  \BibitemShut {NoStop}%
\bibitem [{\citenamefont {Guo}\ \emph {et~al.}(2024{\natexlab{b}})\citenamefont
  {Guo}, \citenamefont {Hart}, \citenamefont {Chen}, \citenamefont {Friedman},\
  and\ \citenamefont {Lucas}}]{guo2024designing}%
  \BibitemOpen
  \bibfield  {author} {\bibinfo {author} {\bibfnamefont {J.}~\bibnamefont
  {Guo}}, \bibinfo {author} {\bibfnamefont {O.}~\bibnamefont {Hart}}, \bibinfo
  {author} {\bibfnamefont {C.-F.}\ \bibnamefont {Chen}}, \bibinfo {author}
  {\bibfnamefont {A.~J.}\ \bibnamefont {Friedman}},\ and\ \bibinfo {author}
  {\bibfnamefont {A.}~\bibnamefont {Lucas}},\ }\href@noop {} {\bibinfo {title}
  {Designing open quantum systems with known steady states: Davies generators
  and beyond}} (\bibinfo {year} {2024}{\natexlab{b}}),\ \Eprint
  {https://arxiv.org/abs/2404.14538} {arXiv:2404.14538 [quant-ph]} \BibitemShut
  {NoStop}%
\bibitem [{\citenamefont {Guo}\ \emph {et~al.}(2024{\natexlab{c}})\citenamefont
  {Guo}, \citenamefont {Ding},\ and\ \citenamefont {Yang}}]{guo2024new}%
  \BibitemOpen
  \bibfield  {author} {\bibinfo {author} {\bibfnamefont {Y.}~\bibnamefont
  {Guo}}, \bibinfo {author} {\bibfnamefont {K.}~\bibnamefont {Ding}},\ and\
  \bibinfo {author} {\bibfnamefont {S.}~\bibnamefont {Yang}},\ }\href
  {https://arxiv.org/abs/2408.03239} {\bibinfo {title} {A new framework for
  quantum phases in open systems: Steady state of imaginary-time lindbladian
  evolution}} (\bibinfo {year} {2024}{\natexlab{c}}),\ \Eprint
  {https://arxiv.org/abs/2408.03239} {arXiv:2408.03239 [quant-ph]} \BibitemShut
  {NoStop}%
\bibitem [{\citenamefont {Ma}\ \emph {et~al.}(2024)\citenamefont {Ma},
  \citenamefont {Zhang}, \citenamefont {Bi}, \citenamefont {Cheng},\ and\
  \citenamefont {Wang}}]{ma2024topological}%
  \BibitemOpen
  \bibfield  {author} {\bibinfo {author} {\bibfnamefont {R.}~\bibnamefont
  {Ma}}, \bibinfo {author} {\bibfnamefont {J.-H.}\ \bibnamefont {Zhang}},
  \bibinfo {author} {\bibfnamefont {Z.}~\bibnamefont {Bi}}, \bibinfo {author}
  {\bibfnamefont {M.}~\bibnamefont {Cheng}},\ and\ \bibinfo {author}
  {\bibfnamefont {C.}~\bibnamefont {Wang}},\ }\href
  {https://arxiv.org/abs/2305.16399} {\bibinfo {title} {Topological phases with
  average symmetries: the decohered, the disordered, and the intrinsic}}
  (\bibinfo {year} {2024}),\ \Eprint {https://arxiv.org/abs/2305.16399}
  {arXiv:2305.16399 [cond-mat.str-el]} \BibitemShut {NoStop}%
\bibitem [{\citenamefont {Lessa}\ \emph
  {et~al.}(2024{\natexlab{b}})\citenamefont {Lessa}, \citenamefont {Ma},
  \citenamefont {Zhang}, \citenamefont {Bi}, \citenamefont {Cheng},\ and\
  \citenamefont {Wang}}]{lessa2024strong}%
  \BibitemOpen
  \bibfield  {author} {\bibinfo {author} {\bibfnamefont {L.~A.}\ \bibnamefont
  {Lessa}}, \bibinfo {author} {\bibfnamefont {R.}~\bibnamefont {Ma}}, \bibinfo
  {author} {\bibfnamefont {J.-H.}\ \bibnamefont {Zhang}}, \bibinfo {author}
  {\bibfnamefont {Z.}~\bibnamefont {Bi}}, \bibinfo {author} {\bibfnamefont
  {M.}~\bibnamefont {Cheng}},\ and\ \bibinfo {author} {\bibfnamefont
  {C.}~\bibnamefont {Wang}},\ }\href {https://arxiv.org/abs/2405.03639}
  {\bibinfo {title} {Strong-to-weak spontaneous symmetry breaking in mixed
  quantum states}} (\bibinfo {year} {2024}{\natexlab{b}}),\ \Eprint
  {https://arxiv.org/abs/2405.03639} {arXiv:2405.03639 [quant-ph]} \BibitemShut
  {NoStop}%
\bibitem [{\citenamefont {Sala}\ \emph
  {et~al.}(2024{\natexlab{a}})\citenamefont {Sala}, \citenamefont
  {Gopalakrishnan}, \citenamefont {Oshikawa},\ and\ \citenamefont
  {You}}]{sala2024spontaneous}%
  \BibitemOpen
  \bibfield  {author} {\bibinfo {author} {\bibfnamefont {P.}~\bibnamefont
  {Sala}}, \bibinfo {author} {\bibfnamefont {S.}~\bibnamefont
  {Gopalakrishnan}}, \bibinfo {author} {\bibfnamefont {M.}~\bibnamefont
  {Oshikawa}},\ and\ \bibinfo {author} {\bibfnamefont {Y.}~\bibnamefont
  {You}},\ }\href {https://arxiv.org/abs/2405.02402} {\bibinfo {title}
  {Spontaneous strong symmetry breaking in open systems: Purification
  perspective}} (\bibinfo {year} {2024}{\natexlab{a}}),\ \Eprint
  {https://arxiv.org/abs/2405.02402} {arXiv:2405.02402 [quant-ph]} \BibitemShut
  {NoStop}%
\bibitem [{\citenamefont {Huang}\ \emph {et~al.}(2024)\citenamefont {Huang},
  \citenamefont {Qi}, \citenamefont {Zhang},\ and\ \citenamefont
  {Lucas}}]{huang2024hydro}%
  \BibitemOpen
  \bibfield  {author} {\bibinfo {author} {\bibfnamefont {X.}~\bibnamefont
  {Huang}}, \bibinfo {author} {\bibfnamefont {M.}~\bibnamefont {Qi}}, \bibinfo
  {author} {\bibfnamefont {J.-H.}\ \bibnamefont {Zhang}},\ and\ \bibinfo
  {author} {\bibfnamefont {A.}~\bibnamefont {Lucas}},\ }\href
  {https://arxiv.org/abs/2407.08760} {\bibinfo {title} {Hydrodynamics as the
  effective field theory of strong-to-weak spontaneous symmetry breaking}}
  (\bibinfo {year} {2024}),\ \Eprint {https://arxiv.org/abs/2407.08760}
  {arXiv:2407.08760 [cond-mat.str-el]} \BibitemShut {NoStop}%
\bibitem [{\citenamefont {Gu}\ \emph {et~al.}(2024)\citenamefont {Gu},
  \citenamefont {Wang},\ and\ \citenamefont {Wang}}]{gu2024spontaneous}%
  \BibitemOpen
  \bibfield  {author} {\bibinfo {author} {\bibfnamefont {D.}~\bibnamefont
  {Gu}}, \bibinfo {author} {\bibfnamefont {Z.}~\bibnamefont {Wang}},\ and\
  \bibinfo {author} {\bibfnamefont {Z.}~\bibnamefont {Wang}},\ }\href
  {https://arxiv.org/abs/2406.19381} {\bibinfo {title} {Spontaneous symmetry
  breaking in open quantum systems: strong, weak, and strong-to-weak}}
  (\bibinfo {year} {2024}),\ \Eprint {https://arxiv.org/abs/2406.19381}
  {arXiv:2406.19381 [quant-ph]} \BibitemShut {NoStop}%
\bibitem [{\citenamefont {Moharramipour}\ \emph {et~al.}(2024)\citenamefont
  {Moharramipour}, \citenamefont {Lessa}, \citenamefont {Wang}, \citenamefont
  {Hsieh},\ and\ \citenamefont {Sahu}}]{moharramipour2024symmetry}%
  \BibitemOpen
  \bibfield  {author} {\bibinfo {author} {\bibfnamefont {A.}~\bibnamefont
  {Moharramipour}}, \bibinfo {author} {\bibfnamefont {L.~A.}\ \bibnamefont
  {Lessa}}, \bibinfo {author} {\bibfnamefont {C.}~\bibnamefont {Wang}},
  \bibinfo {author} {\bibfnamefont {T.~H.}\ \bibnamefont {Hsieh}},\ and\
  \bibinfo {author} {\bibfnamefont {S.}~\bibnamefont {Sahu}},\ }\href
  {https://arxiv.org/abs/2406.08542} {\bibinfo {title} {Symmetry enforced
  entanglement in maximally mixed states}} (\bibinfo {year} {2024}),\ \Eprint
  {https://arxiv.org/abs/2406.08542} {arXiv:2406.08542 [quant-ph]} \BibitemShut
  {NoStop}%
\bibitem [{\citenamefont {Kuno}\ \emph {et~al.}(2024)\citenamefont {Kuno},
  \citenamefont {Orito},\ and\ \citenamefont {Ichinose}}]{kuno2024strong}%
  \BibitemOpen
  \bibfield  {author} {\bibinfo {author} {\bibfnamefont {Y.}~\bibnamefont
  {Kuno}}, \bibinfo {author} {\bibfnamefont {T.}~\bibnamefont {Orito}},\ and\
  \bibinfo {author} {\bibfnamefont {I.}~\bibnamefont {Ichinose}},\ }\href
  {https://arxiv.org/abs/2408.04241} {\bibinfo {title} {Strong-to-weak symmetry
  breaking states in stochastic dephasing stabilizer circuits}} (\bibinfo
  {year} {2024}),\ \Eprint {https://arxiv.org/abs/2408.04241} {arXiv:2408.04241
  [quant-ph]} \BibitemShut {NoStop}%
\bibitem [{\citenamefont {Su}\ \emph {et~al.}(2024)\citenamefont {Su},
  \citenamefont {Bao},\ and\ \citenamefont {Xu}}]{su2024emergent}%
  \BibitemOpen
  \bibfield  {author} {\bibinfo {author} {\bibfnamefont {K.}~\bibnamefont
  {Su}}, \bibinfo {author} {\bibfnamefont {Y.}~\bibnamefont {Bao}},\ and\
  \bibinfo {author} {\bibfnamefont {C.}~\bibnamefont {Xu}},\ }\href
  {https://arxiv.org/abs/2408.07125} {\bibinfo {title} {Emergent gauge fields
  and the "choi-spin liquids" in steady states}} (\bibinfo {year} {2024}),\
  \Eprint {https://arxiv.org/abs/2408.07125} {arXiv:2408.07125
  [cond-mat.str-el]} \BibitemShut {NoStop}%
\bibitem [{\citenamefont {Sala}\ \emph
  {et~al.}(2024{\natexlab{b}})\citenamefont {Sala}, \citenamefont {Alicea},\
  and\ \citenamefont {Verresen}}]{sala2024decoherence}%
  \BibitemOpen
  \bibfield  {author} {\bibinfo {author} {\bibfnamefont {P.}~\bibnamefont
  {Sala}}, \bibinfo {author} {\bibfnamefont {J.}~\bibnamefont {Alicea}},\ and\
  \bibinfo {author} {\bibfnamefont {R.}~\bibnamefont {Verresen}},\ }\href
  {https://arxiv.org/abs/2409.12948} {\bibinfo {title} {Decoherence and
  wavefunction deformation of $d_4$ non-abelian topological order}} (\bibinfo
  {year} {2024}{\natexlab{b}}),\ \Eprint {https://arxiv.org/abs/2409.12948}
  {arXiv:2409.12948 [cond-mat.str-el]} \BibitemShut {NoStop}%
\bibitem [{\citenamefont {Zhang}\ \emph {et~al.}(2024)\citenamefont {Zhang},
  \citenamefont {Xu}, \citenamefont {Zhang}, \citenamefont {Xu}, \citenamefont
  {Bi},\ and\ \citenamefont {Luo}}]{zhang2024strong}%
  \BibitemOpen
  \bibfield  {author} {\bibinfo {author} {\bibfnamefont {C.}~\bibnamefont
  {Zhang}}, \bibinfo {author} {\bibfnamefont {Y.}~\bibnamefont {Xu}}, \bibinfo
  {author} {\bibfnamefont {J.-H.}\ \bibnamefont {Zhang}}, \bibinfo {author}
  {\bibfnamefont {C.}~\bibnamefont {Xu}}, \bibinfo {author} {\bibfnamefont
  {Z.}~\bibnamefont {Bi}},\ and\ \bibinfo {author} {\bibfnamefont {Z.-X.}\
  \bibnamefont {Luo}},\ }\href {https://arxiv.org/abs/2409.17530} {\bibinfo
  {title} {Strong-to-weak spontaneous breaking of 1-form symmetry and
  intrinsically mixed topological order}} (\bibinfo {year} {2024}),\ \Eprint
  {https://arxiv.org/abs/2409.17530} {arXiv:2409.17530 [quant-ph]} \BibitemShut
  {NoStop}%
\bibitem [{\citenamefont {Sachdev}(2011)}]{Sachdev_2011}%
  \BibitemOpen
  \bibfield  {author} {\bibinfo {author} {\bibfnamefont {S.}~\bibnamefont
  {Sachdev}},\ }\href@noop {} {\emph {\bibinfo {title} {Quantum Phase
  Transitions}}},\ \bibinfo {edition} {2nd}\ ed.\ (\bibinfo  {publisher}
  {Cambridge University Press},\ \bibinfo {year} {2011})\BibitemShut {NoStop}%
\bibitem [{\citenamefont {Mahan}(2013)}]{mahan2013many}%
  \BibitemOpen
  \bibfield  {author} {\bibinfo {author} {\bibfnamefont {G.~D.}\ \bibnamefont
  {Mahan}},\ }\href@noop {} {\emph {\bibinfo {title} {Many-particle physics}}}\
  (\bibinfo  {publisher} {Springer Science \& Business Media},\ \bibinfo {year}
  {2013})\BibitemShut {NoStop}%
\bibitem [{\citenamefont {Callen}\ and\ \citenamefont
  {Welton}(1951)}]{callen1951irreversibility}%
  \BibitemOpen
  \bibfield  {author} {\bibinfo {author} {\bibfnamefont {H.~B.}\ \bibnamefont
  {Callen}}\ and\ \bibinfo {author} {\bibfnamefont {T.~A.}\ \bibnamefont
  {Welton}},\ }\bibfield  {title} {\bibinfo {title} {Irreversibility and
  generalized noise},\ }\href@noop {} {\bibfield  {journal} {\bibinfo
  {journal} {Physical Review}\ }\textbf {\bibinfo {volume} {83}},\ \bibinfo
  {pages} {34} (\bibinfo {year} {1951})}\BibitemShut {NoStop}%
\bibitem [{\citenamefont {Kubo}(1966)}]{kubo1966fluctuation}%
  \BibitemOpen
  \bibfield  {author} {\bibinfo {author} {\bibfnamefont {R.}~\bibnamefont
  {Kubo}},\ }\bibfield  {title} {\bibinfo {title} {The fluctuation-dissipation
  theorem},\ }\href@noop {} {\bibfield  {journal} {\bibinfo  {journal} {Reports
  on progress in physics}\ }\textbf {\bibinfo {volume} {29}},\ \bibinfo {pages}
  {255} (\bibinfo {year} {1966})}\BibitemShut {NoStop}%
\bibitem [{\citenamefont {Marconi}\ \emph {et~al.}(2008)\citenamefont
  {Marconi}, \citenamefont {Puglisi}, \citenamefont {Rondoni},\ and\
  \citenamefont {Vulpiani}}]{marconi2008fluctuation}%
  \BibitemOpen
  \bibfield  {author} {\bibinfo {author} {\bibfnamefont {U.~M.~B.}\
  \bibnamefont {Marconi}}, \bibinfo {author} {\bibfnamefont {A.}~\bibnamefont
  {Puglisi}}, \bibinfo {author} {\bibfnamefont {L.}~\bibnamefont {Rondoni}},\
  and\ \bibinfo {author} {\bibfnamefont {A.}~\bibnamefont {Vulpiani}},\
  }\bibfield  {title} {\bibinfo {title} {Fluctuation--dissipation: response
  theory in statistical physics},\ }\href@noop {} {\bibfield  {journal}
  {\bibinfo  {journal} {Physics reports}\ }\textbf {\bibinfo {volume} {461}},\
  \bibinfo {pages} {111} (\bibinfo {year} {2008})}\BibitemShut {NoStop}%
\bibitem [{\citenamefont {Cozzini}\ \emph {et~al.}(2007)\citenamefont
  {Cozzini}, \citenamefont {Ionicioiu},\ and\ \citenamefont
  {Zanardi}}]{cozzini2007quantum}%
  \BibitemOpen
  \bibfield  {author} {\bibinfo {author} {\bibfnamefont {M.}~\bibnamefont
  {Cozzini}}, \bibinfo {author} {\bibfnamefont {R.}~\bibnamefont {Ionicioiu}},\
  and\ \bibinfo {author} {\bibfnamefont {P.}~\bibnamefont {Zanardi}},\
  }\bibfield  {title} {\bibinfo {title} {Quantum fidelity and quantum phase
  transitions in matrix product states},\ }\href@noop {} {\bibfield  {journal}
  {\bibinfo  {journal} {Physical Review B—Condensed Matter and Materials
  Physics}\ }\textbf {\bibinfo {volume} {76}},\ \bibinfo {pages} {104420}
  (\bibinfo {year} {2007})}\BibitemShut {NoStop}%
\bibitem [{\citenamefont {You}\ \emph {et~al.}(2007)\citenamefont {You},
  \citenamefont {Li},\ and\ \citenamefont {Gu}}]{you2007fidelity}%
  \BibitemOpen
  \bibfield  {author} {\bibinfo {author} {\bibfnamefont {W.-L.}\ \bibnamefont
  {You}}, \bibinfo {author} {\bibfnamefont {Y.-W.}\ \bibnamefont {Li}},\ and\
  \bibinfo {author} {\bibfnamefont {S.-J.}\ \bibnamefont {Gu}},\ }\bibfield
  {title} {\bibinfo {title} {Fidelity, dynamic structure factor, and
  susceptibility in critical phenomena},\ }\href@noop {} {\bibfield  {journal}
  {\bibinfo  {journal} {Physical Review E—Statistical, Nonlinear, and Soft
  Matter Physics}\ }\textbf {\bibinfo {volume} {76}},\ \bibinfo {pages}
  {022101} (\bibinfo {year} {2007})}\BibitemShut {NoStop}%
\bibitem [{\citenamefont {Chen}\ \emph {et~al.}(2008)\citenamefont {Chen},
  \citenamefont {Wang}, \citenamefont {Hao},\ and\ \citenamefont
  {Wang}}]{chen2008intrinsic}%
  \BibitemOpen
  \bibfield  {author} {\bibinfo {author} {\bibfnamefont {S.}~\bibnamefont
  {Chen}}, \bibinfo {author} {\bibfnamefont {L.}~\bibnamefont {Wang}}, \bibinfo
  {author} {\bibfnamefont {Y.}~\bibnamefont {Hao}},\ and\ \bibinfo {author}
  {\bibfnamefont {Y.}~\bibnamefont {Wang}},\ }\bibfield  {title} {\bibinfo
  {title} {Intrinsic relation between ground-state fidelity and the
  characterization of a quantum phase transition},\ }\href@noop {} {\bibfield
  {journal} {\bibinfo  {journal} {Physical Review A—Atomic, Molecular, and
  Optical Physics}\ }\textbf {\bibinfo {volume} {77}},\ \bibinfo {pages}
  {032111} (\bibinfo {year} {2008})}\BibitemShut {NoStop}%
\bibitem [{\citenamefont {Zhou}\ and\ \citenamefont
  {Barjaktarevi{\v{c}}}(2008)}]{zhou2008fidelity}%
  \BibitemOpen
  \bibfield  {author} {\bibinfo {author} {\bibfnamefont {H.-Q.}\ \bibnamefont
  {Zhou}}\ and\ \bibinfo {author} {\bibfnamefont {J.~P.}\ \bibnamefont
  {Barjaktarevi{\v{c}}}},\ }\bibfield  {title} {\bibinfo {title} {Fidelity and
  quantum phase transitions},\ }\href@noop {} {\bibfield  {journal} {\bibinfo
  {journal} {Journal of Physics A: Mathematical and Theoretical}\ }\textbf
  {\bibinfo {volume} {41}},\ \bibinfo {pages} {412001} (\bibinfo {year}
  {2008})}\BibitemShut {NoStop}%
\bibitem [{\citenamefont {Quan}\ and\ \citenamefont
  {Cucchietti}(2009)}]{quan2009quantum}%
  \BibitemOpen
  \bibfield  {author} {\bibinfo {author} {\bibfnamefont {H.}~\bibnamefont
  {Quan}}\ and\ \bibinfo {author} {\bibfnamefont {F.}~\bibnamefont
  {Cucchietti}},\ }\bibfield  {title} {\bibinfo {title} {Quantum fidelity and
  thermal phase transitions},\ }\href@noop {} {\bibfield  {journal} {\bibinfo
  {journal} {Physical Review E—Statistical, Nonlinear, and Soft Matter
  Physics}\ }\textbf {\bibinfo {volume} {79}},\ \bibinfo {pages} {031101}
  (\bibinfo {year} {2009})}\BibitemShut {NoStop}%
\bibitem [{\citenamefont {Gu}(2010)}]{gu2010fidelity}%
  \BibitemOpen
  \bibfield  {author} {\bibinfo {author} {\bibfnamefont {S.-J.}\ \bibnamefont
  {Gu}},\ }\bibfield  {title} {\bibinfo {title} {Fidelity approach to quantum
  phase transitions},\ }\href@noop {} {\bibfield  {journal} {\bibinfo
  {journal} {International Journal of Modern Physics B}\ }\textbf {\bibinfo
  {volume} {24}},\ \bibinfo {pages} {4371} (\bibinfo {year}
  {2010})}\BibitemShut {NoStop}%
\bibitem [{\citenamefont {Albuquerque}\ \emph {et~al.}(2010)\citenamefont
  {Albuquerque}, \citenamefont {Alet}, \citenamefont {Sire},\ and\
  \citenamefont {Capponi}}]{albuquerque2010quantum}%
  \BibitemOpen
  \bibfield  {author} {\bibinfo {author} {\bibfnamefont {A.~F.}\ \bibnamefont
  {Albuquerque}}, \bibinfo {author} {\bibfnamefont {F.}~\bibnamefont {Alet}},
  \bibinfo {author} {\bibfnamefont {C.}~\bibnamefont {Sire}},\ and\ \bibinfo
  {author} {\bibfnamefont {S.}~\bibnamefont {Capponi}},\ }\bibfield  {title}
  {\bibinfo {title} {Quantum critical scaling of fidelity susceptibility},\
  }\href@noop {} {\bibfield  {journal} {\bibinfo  {journal} {Physical Review
  B—Condensed Matter and Materials Physics}\ }\textbf {\bibinfo {volume}
  {81}},\ \bibinfo {pages} {064418} (\bibinfo {year} {2010})}\BibitemShut
  {NoStop}%
\bibitem [{\citenamefont {Banchi}\ \emph {et~al.}(2014)\citenamefont {Banchi},
  \citenamefont {Giorda},\ and\ \citenamefont {Zanardi}}]{banchi2014quantum}%
  \BibitemOpen
  \bibfield  {author} {\bibinfo {author} {\bibfnamefont {L.}~\bibnamefont
  {Banchi}}, \bibinfo {author} {\bibfnamefont {P.}~\bibnamefont {Giorda}},\
  and\ \bibinfo {author} {\bibfnamefont {P.}~\bibnamefont {Zanardi}},\
  }\bibfield  {title} {\bibinfo {title} {Quantum information-geometry of
  dissipative quantum phase transitions},\ }\href@noop {} {\bibfield  {journal}
  {\bibinfo  {journal} {Physical Review E}\ }\textbf {\bibinfo {volume} {89}},\
  \bibinfo {pages} {022102} (\bibinfo {year} {2014})}\BibitemShut {NoStop}%
\bibitem [{\citenamefont {Carollo}\ \emph {et~al.}(2018)\citenamefont
  {Carollo}, \citenamefont {Spagnolo},\ and\ \citenamefont
  {Valenti}}]{carollo2018uhlmann}%
  \BibitemOpen
  \bibfield  {author} {\bibinfo {author} {\bibfnamefont {A.}~\bibnamefont
  {Carollo}}, \bibinfo {author} {\bibfnamefont {B.}~\bibnamefont {Spagnolo}},\
  and\ \bibinfo {author} {\bibfnamefont {D.}~\bibnamefont {Valenti}},\
  }\bibfield  {title} {\bibinfo {title} {Uhlmann curvature in dissipative phase
  transitions},\ }\href@noop {} {\bibfield  {journal} {\bibinfo  {journal}
  {Scientific reports}\ }\textbf {\bibinfo {volume} {8}},\ \bibinfo {pages}
  {9852} (\bibinfo {year} {2018})}\BibitemShut {NoStop}%
\bibitem [{\citenamefont {Carollo}\ \emph {et~al.}(2020)\citenamefont
  {Carollo}, \citenamefont {Valenti},\ and\ \citenamefont
  {Spagnolo}}]{carollo2020geometry}%
  \BibitemOpen
  \bibfield  {author} {\bibinfo {author} {\bibfnamefont {A.}~\bibnamefont
  {Carollo}}, \bibinfo {author} {\bibfnamefont {D.}~\bibnamefont {Valenti}},\
  and\ \bibinfo {author} {\bibfnamefont {B.}~\bibnamefont {Spagnolo}},\
  }\bibfield  {title} {\bibinfo {title} {Geometry of quantum phase
  transitions},\ }\href@noop {} {\bibfield  {journal} {\bibinfo  {journal}
  {Physics Reports}\ }\textbf {\bibinfo {volume} {838}},\ \bibinfo {pages} {1}
  (\bibinfo {year} {2020})}\BibitemShut {NoStop}%
\bibitem [{sup()}]{supplementary}%
  \BibitemOpen
  \href@noop {} {\bibinfo  {journal} {see Supplementary Materials for more
  details}\ }\BibitemShut {NoStop}%
\bibitem [{\citenamefont {Bures}(1969)}]{bures1969extension}%
  \BibitemOpen
\bibfield  {journal} {  }\bibfield  {author} {\bibinfo {author} {\bibfnamefont
  {D.}~\bibnamefont {Bures}},\ }\bibfield  {title} {\bibinfo {title} {An
  extension of kakutani's theorem on infinite product measures to the tensor
  product of semifinite w*-algebras},\ }\href@noop {} {\bibfield  {journal}
  {\bibinfo  {journal} {Transactions of the American Mathematical Society}\
  }\textbf {\bibinfo {volume} {135}},\ \bibinfo {pages} {199} (\bibinfo {year}
  {1969})}\BibitemShut {NoStop}%
\bibitem [{\citenamefont {Uhlmann}(1976)}]{Uhlmann}%
  \BibitemOpen
  \bibfield  {author} {\bibinfo {author} {\bibfnamefont {A.}~\bibnamefont
  {Uhlmann}},\ }\bibfield  {title} {\bibinfo {title} {The “transition
  probability” in the state space of a*-algebra},\ }\href
  {https://doi.org/https://doi.org/10.1016/0034-4877(76)90060-4} {\bibfield
  {journal} {\bibinfo  {journal} {Reports on Mathematical Physics}\ }\textbf
  {\bibinfo {volume} {9}},\ \bibinfo {pages} {273} (\bibinfo {year}
  {1976})}\BibitemShut {NoStop}%
\bibitem [{\citenamefont {H{\"u}bner}(1992)}]{hubner1992explicit}%
  \BibitemOpen
  \bibfield  {author} {\bibinfo {author} {\bibfnamefont {M.}~\bibnamefont
  {H{\"u}bner}},\ }\bibfield  {title} {\bibinfo {title} {Explicit computation
  of the bures distance for density matrices},\ }\href@noop {} {\bibfield
  {journal} {\bibinfo  {journal} {Physics Letters A}\ }\textbf {\bibinfo
  {volume} {163}},\ \bibinfo {pages} {239} (\bibinfo {year}
  {1992})}\BibitemShut {NoStop}%
\bibitem [{\citenamefont {Braunstein}\ and\ \citenamefont
  {Caves}(1994)}]{braunstein1994statistical}%
  \BibitemOpen
  \bibfield  {author} {\bibinfo {author} {\bibfnamefont {S.~L.}\ \bibnamefont
  {Braunstein}}\ and\ \bibinfo {author} {\bibfnamefont {C.~M.}\ \bibnamefont
  {Caves}},\ }\bibfield  {title} {\bibinfo {title} {Statistical distance and
  the geometry of quantum states},\ }\href@noop {} {\bibfield  {journal}
  {\bibinfo  {journal} {Physical Review Letters}\ }\textbf {\bibinfo {volume}
  {72}},\ \bibinfo {pages} {3439} (\bibinfo {year} {1994})}\BibitemShut
  {NoStop}%
\bibitem [{\citenamefont {Paris}(2009)}]{paris2009quantum}%
  \BibitemOpen
  \bibfield  {author} {\bibinfo {author} {\bibfnamefont {M.~G.}\ \bibnamefont
  {Paris}},\ }\bibfield  {title} {\bibinfo {title} {Quantum estimation for
  quantum technology},\ }\href@noop {} {\bibfield  {journal} {\bibinfo
  {journal} {International Journal of Quantum Information}\ }\textbf {\bibinfo
  {volume} {7}},\ \bibinfo {pages} {125} (\bibinfo {year} {2009})}\BibitemShut
  {NoStop}%
\bibitem [{\citenamefont {Holevo}(2011)}]{holevo2011probabilistic}%
  \BibitemOpen
  \bibfield  {author} {\bibinfo {author} {\bibfnamefont {A.~S.}\ \bibnamefont
  {Holevo}},\ }\href@noop {} {\emph {\bibinfo {title} {Probabilistic and
  statistical aspects of quantum theory}}},\ Vol.~\bibinfo {volume} {1}\
  (\bibinfo  {publisher} {Springer Science \& Business Media},\ \bibinfo {year}
  {2011})\BibitemShut {NoStop}%
\bibitem [{\citenamefont {Bengtsson}\ and\ \citenamefont
  {{\.Z}yczkowski}(2017)}]{bengtsson2017geometry}%
  \BibitemOpen
  \bibfield  {author} {\bibinfo {author} {\bibfnamefont {I.}~\bibnamefont
  {Bengtsson}}\ and\ \bibinfo {author} {\bibfnamefont {K.}~\bibnamefont
  {{\.Z}yczkowski}},\ }\href@noop {} {\emph {\bibinfo {title} {Geometry of
  quantum states: an introduction to quantum entanglement}}}\ (\bibinfo
  {publisher} {Cambridge university press},\ \bibinfo {year}
  {2017})\BibitemShut {NoStop}%
\bibitem [{\citenamefont {Grace}\ and\ \citenamefont {Guha}(2022)}]{9965836}%
  \BibitemOpen
  \bibfield  {author} {\bibinfo {author} {\bibfnamefont {M.~R.}\ \bibnamefont
  {Grace}}\ and\ \bibinfo {author} {\bibfnamefont {S.}~\bibnamefont {Guha}},\
  }\bibfield  {title} {\bibinfo {title} {Perturbation theory for quantum
  information},\ }in\ \href {https://doi.org/10.1109/ITW54588.2022.9965836}
  {\emph {\bibinfo {booktitle} {2022 IEEE Information Theory Workshop (ITW)}}}\
  (\bibinfo {year} {2022})\ pp.\ \bibinfo {pages} {500--505}\BibitemShut
  {NoStop}%
\bibitem [{\citenamefont
  {{\v{S}}afr{\'a}nek}(2017)}]{vsafranek2017discontinuities}%
  \BibitemOpen
  \bibfield  {author} {\bibinfo {author} {\bibfnamefont {D.}~\bibnamefont
  {{\v{S}}afr{\'a}nek}},\ }\bibfield  {title} {\bibinfo {title}
  {Discontinuities of the quantum fisher information and the bures metric},\
  }\href@noop {} {\bibfield  {journal} {\bibinfo  {journal} {Physical Review
  A}\ }\textbf {\bibinfo {volume} {95}},\ \bibinfo {pages} {052320} (\bibinfo
  {year} {2017})}\BibitemShut {NoStop}%
\bibitem [{\citenamefont {Zhou}\ and\ \citenamefont
  {Jiang}(2019)}]{zhou2019exact}%
  \BibitemOpen
  \bibfield  {author} {\bibinfo {author} {\bibfnamefont {S.}~\bibnamefont
  {Zhou}}\ and\ \bibinfo {author} {\bibfnamefont {L.}~\bibnamefont {Jiang}},\
  }\href {https://arxiv.org/abs/1910.08473} {\bibinfo {title} {An exact
  correspondence between the quantum fisher information and the bures metric}}
  (\bibinfo {year} {2019}),\ \Eprint {https://arxiv.org/abs/1910.08473}
  {arXiv:1910.08473 [quant-ph]} \BibitemShut {NoStop}%
\bibitem [{\citenamefont {Meyer}(2021)}]{meyer2021fisher}%
  \BibitemOpen
  \bibfield  {author} {\bibinfo {author} {\bibfnamefont {J.~J.}\ \bibnamefont
  {Meyer}},\ }\bibfield  {title} {\bibinfo {title} {Fisher information in noisy
  intermediate-scale quantum applications},\ }\href@noop {} {\bibfield
  {journal} {\bibinfo  {journal} {Quantum}\ }\textbf {\bibinfo {volume} {5}},\
  \bibinfo {pages} {539} (\bibinfo {year} {2021})}\BibitemShut {NoStop}%
\bibitem [{\citenamefont {Anderson}(1967)}]{anderson}%
  \BibitemOpen
  \bibfield  {author} {\bibinfo {author} {\bibfnamefont {P.~W.}\ \bibnamefont
  {Anderson}},\ }\bibfield  {title} {\bibinfo {title} {Infrared catastrophe in
  fermi gases with local scattering potentials},\ }\href
  {https://doi.org/10.1103/PhysRevLett.18.1049} {\bibfield  {journal} {\bibinfo
   {journal} {Phys. Rev. Lett.}\ }\textbf {\bibinfo {volume} {18}},\ \bibinfo
  {pages} {1049} (\bibinfo {year} {1967})}\BibitemShut {NoStop}%
\bibitem [{\citenamefont {Miszczak}\ \emph {et~al.}(2008)\citenamefont
  {Miszczak}, \citenamefont {Pucha{\l}a}, \citenamefont {Horodecki},
  \citenamefont {Uhlmann},\ and\ \citenamefont
  {{\.Z}yczkowski}}]{miszczak2008sub}%
  \BibitemOpen
  \bibfield  {author} {\bibinfo {author} {\bibfnamefont {J.~A.}\ \bibnamefont
  {Miszczak}}, \bibinfo {author} {\bibfnamefont {Z.}~\bibnamefont
  {Pucha{\l}a}}, \bibinfo {author} {\bibfnamefont {P.}~\bibnamefont
  {Horodecki}}, \bibinfo {author} {\bibfnamefont {A.}~\bibnamefont {Uhlmann}},\
  and\ \bibinfo {author} {\bibfnamefont {K.}~\bibnamefont {{\.Z}yczkowski}},\
  }\bibfield  {title} {\bibinfo {title} {Sub--and super--fidelity as bounds for
  quantum fidelity},\ }\href@noop {} {\bibfield  {journal} {\bibinfo  {journal}
  {arXiv preprint arXiv:0805.2037}\ } (\bibinfo {year} {2008})}\BibitemShut
  {NoStop}%
\bibitem [{\citenamefont {Huang}\ \emph {et~al.}(2020)\citenamefont {Huang},
  \citenamefont {Kueng},\ and\ \citenamefont {Preskill}}]{Huang_2020}%
  \BibitemOpen
  \bibfield  {author} {\bibinfo {author} {\bibfnamefont {H.-Y.}\ \bibnamefont
  {Huang}}, \bibinfo {author} {\bibfnamefont {R.}~\bibnamefont {Kueng}},\ and\
  \bibinfo {author} {\bibfnamefont {J.}~\bibnamefont {Preskill}},\ }\bibfield
  {title} {\bibinfo {title} {Predicting many properties of a quantum system
  from very few measurements},\ }\href
  {https://doi.org/10.1038/s41567-020-0932-7} {\bibfield  {journal} {\bibinfo
  {journal} {Nature Physics}\ }\textbf {\bibinfo {volume} {16}},\ \bibinfo
  {pages} {1050–1057} (\bibinfo {year} {2020})}\BibitemShut {NoStop}%
\bibitem [{\citenamefont {Elben}\ \emph {et~al.}(2022)\citenamefont {Elben},
  \citenamefont {Flammia}, \citenamefont {Huang}, \citenamefont {Kueng},
  \citenamefont {Preskill}, \citenamefont {Vermersch},\ and\ \citenamefont
  {Zoller}}]{Elben_2022}%
  \BibitemOpen
  \bibfield  {author} {\bibinfo {author} {\bibfnamefont {A.}~\bibnamefont
  {Elben}}, \bibinfo {author} {\bibfnamefont {S.~T.}\ \bibnamefont {Flammia}},
  \bibinfo {author} {\bibfnamefont {H.-Y.}\ \bibnamefont {Huang}}, \bibinfo
  {author} {\bibfnamefont {R.}~\bibnamefont {Kueng}}, \bibinfo {author}
  {\bibfnamefont {J.}~\bibnamefont {Preskill}}, \bibinfo {author}
  {\bibfnamefont {B.}~\bibnamefont {Vermersch}},\ and\ \bibinfo {author}
  {\bibfnamefont {P.}~\bibnamefont {Zoller}},\ }\bibfield  {title} {\bibinfo
  {title} {The randomized measurement toolbox},\ }\href
  {https://doi.org/10.1038/s42254-022-00535-2} {\bibfield  {journal} {\bibinfo
  {journal} {Nature Reviews Physics}\ }\textbf {\bibinfo {volume} {5}},\
  \bibinfo {pages} {9–24} (\bibinfo {year} {2022})}\BibitemShut {NoStop}%
\bibitem [{\citenamefont {Hu}\ and\ \citenamefont
  {You}(2022)}]{hu2022Hamiltonian}%
  \BibitemOpen
  \bibfield  {author} {\bibinfo {author} {\bibfnamefont {H.-Y.}\ \bibnamefont
  {Hu}}\ and\ \bibinfo {author} {\bibfnamefont {Y.-Z.}\ \bibnamefont {You}},\
  }\bibfield  {title} {\bibinfo {title} {Hamiltonian-driven shadow tomography
  of quantum states},\ }\href
  {https://doi.org/10.1103/PhysRevResearch.4.013054} {\bibfield  {journal}
  {\bibinfo  {journal} {Phys. Rev. Res.}\ }\textbf {\bibinfo {volume} {4}},\
  \bibinfo {pages} {013054} (\bibinfo {year} {2022})}\BibitemShut {NoStop}%
\bibitem [{\citenamefont {Hu}\ \emph {et~al.}(2023)\citenamefont {Hu},
  \citenamefont {Choi},\ and\ \citenamefont {You}}]{hu2023classical}%
  \BibitemOpen
  \bibfield  {author} {\bibinfo {author} {\bibfnamefont {H.-Y.}\ \bibnamefont
  {Hu}}, \bibinfo {author} {\bibfnamefont {S.}~\bibnamefont {Choi}},\ and\
  \bibinfo {author} {\bibfnamefont {Y.-Z.}\ \bibnamefont {You}},\ }\bibfield
  {title} {\bibinfo {title} {Classical shadow tomography with locally scrambled
  quantum dynamics},\ }\href {https://doi.org/10.1103/PhysRevResearch.5.023027}
  {\bibfield  {journal} {\bibinfo  {journal} {Phys. Rev. Res.}\ }\textbf
  {\bibinfo {volume} {5}},\ \bibinfo {pages} {023027} (\bibinfo {year}
  {2023})}\BibitemShut {NoStop}%
\bibitem [{\citenamefont {Ando}(1988)}]{ando1988comparison}%
  \BibitemOpen
  \bibfield  {author} {\bibinfo {author} {\bibfnamefont {T.}~\bibnamefont
  {Ando}},\ }\bibfield  {title} {\bibinfo {title} {Comparison of norms $||| f
  (a)- f (b)|||$ and $||| f (| a- b|)|||$},\ }\href@noop {} {\bibfield
  {journal} {\bibinfo  {journal} {Mathematische Zeitschrift}\ }\textbf
  {\bibinfo {volume} {197}},\ \bibinfo {pages} {403} (\bibinfo {year}
  {1988})}\BibitemShut {NoStop}%
\bibitem [{\citenamefont {Ando}\ and\ \citenamefont
  {Zhan}(1999)}]{ando1999norm}%
  \BibitemOpen
  \bibfield  {author} {\bibinfo {author} {\bibfnamefont {T.}~\bibnamefont
  {Ando}}\ and\ \bibinfo {author} {\bibfnamefont {X.}~\bibnamefont {Zhan}},\
  }\bibfield  {title} {\bibinfo {title} {Norm inequalities related to operator
  monotone functions},\ }\href@noop {} {\bibfield  {journal} {\bibinfo
  {journal} {Mathematische Annalen}\ }\textbf {\bibinfo {volume} {315}},\
  \bibinfo {pages} {771} (\bibinfo {year} {1999})}\BibitemShut {NoStop}%
\bibitem [{\citenamefont {Baldwin}\ and\ \citenamefont
  {Jones}(2023)}]{baldwin2023efficiently}%
  \BibitemOpen
  \bibfield  {author} {\bibinfo {author} {\bibfnamefont {A.~J.}\ \bibnamefont
  {Baldwin}}\ and\ \bibinfo {author} {\bibfnamefont {J.~A.}\ \bibnamefont
  {Jones}},\ }\bibfield  {title} {\bibinfo {title} {Efficiently computing the
  uhlmann fidelity for density matrices},\ }\href@noop {} {\bibfield  {journal}
  {\bibinfo  {journal} {Physical Review A}\ }\textbf {\bibinfo {volume}
  {107}},\ \bibinfo {pages} {012427} (\bibinfo {year} {2023})}\BibitemShut
  {NoStop}%
\end{thebibliography}%

\appendix
\setcounter{equation}{0}
\renewcommand{\thesection}{S-\arabic{section}} \renewcommand{\theequation}{S%
\arabic{equation}} \setcounter{equation}{0} \renewcommand{\thefigure}{S%
\arabic{figure}} \setcounter{figure}{0}

\onecolumngrid
\newpage

\vskip0.2cm
\centerline{\large\textbf{Supplemental Materials for ``Fluctuation-Dissipation Theorem and Information Geometry in}}
\vskip0.12cm
\centerline{\large\textbf{Open Quantum Systems''}}

\vskip0.8cm

\maketitle

\section{\rom{1}. Properties of fidelity and distances}
We summarize some useful properties of fidelity, Bures distance and trace distance that are used in the main text and the supplementary material.
\begin{enumerate}[1.]

\item Additivity: for density matrices $\rho_i$, $\sigma_i$ and parameters $\lambda_i, \mu_i\in[0,1]$ ($i=1,\cdots, N$),
\begin{align}\label{eq: additivity}
\begin{gathered}
\sqrt{F}\left(\bigoplus_{i=1}^N\lambda_i\rho_i,\bigoplus_{i=1}^N\mu_i\sigma_i\right)=\sum_{i=1}^N\sqrt{\lambda_i\mu_i}\sqrt{F}(\rho_i, \sigma_i).
\end{gathered}
\end{align}
In particular, with the presence of strong global symmetry, the Hilbert space can be decomposed into a direct sum of subspaces with different charges for the global symmetry. Therefore, the additivity applies to summations of density matrices over different global symmetry charges, since the algebraic sum is now equivalent to the direct sum.

\item Concavity: for density matrices $\rho$ and $\sigma_i$ ($i=1,\cdots, N$),
\begin{align}
\begin{gathered}
F\left(\rho, \sum_ip_i\sigma_i\right) \geq \sum_ip_iF(\rho, \sigma_i)
\end{gathered},
\end{align}
where $\{p_i\}$ is a probability distribution such that $\sum_ip_i=1$.
\item Joint concavity: for density matrices $\rho_i$ and $\sigma_i$ ($i=1,\cdots, N$) with the probability distribution $\{p_i\}$,
\begin{align}
\begin{gathered}
\sqrt{F}\left(\sum_ip_i\rho_i,\sum_ip_i\sigma_i\right) \geq \sum_ip_i\sqrt{F}(\rho_i,\sigma_i)
\end{gathered}.
\end{align}
\item Multiplicativity: for density matrices $\rho_i$ and $\sigma_i$ ($i=1,2$),
\begin{align}
\begin{gathered}
F(\rho_1\otimes\rho_2,\sigma_1\otimes\sigma_2) = F(\rho_1, \sigma_1)F(\rho_2, \sigma_2)
\end{gathered}.
\end{align}
\item Data processing inequality: for a completely positive trace-preserving map $\E$ (i.e., quantum channel),
\begin{align}\label{eq: dataprocess}
\begin{gathered}
F\left(\E[\rho], \E[\sigma]\right) \geq F(\rho, \sigma)
\end{gathered}.
\end{align}

\end{enumerate}

\section{\rom{2}. Upper bound of the fidelity susceptibility in Eq. \eqref{eq: chifinequal}}
In this section, we prove that the upper bound in Eq. \eqref{eq: chifinequal}. To this end, we first note that, for positive definite matrices $A$ and $B$ of the same dimension, the following inequality holds:
\begin{equation}\label{eq: lemma}
    \tr\sqrt{A+B}\leq\tr\sqrt{A}+\tr\sqrt{B}.
\end{equation}
This is known as Ando's inequality in the context of operator monotone functions \cite{ando1988comparison,ando1999norm}.

Applying Eq. \eqref{eq: lemma} recursively, we arrive at the following inequality: for positive semi-definite matrices $M_i$ of the same dimension, $i=1,2,\dots, N$, we have
\begin{equation}\label{eq: lemma2}
    \tr\left(\sqrt{\sum_{i=1}^N M_i}\right)\leq \sum_{i=1}^N\tr\sqrt{M_i}.
\end{equation}

Now consider the fidelity susceptibility. Assuming translation invariance, we have
\begin{align}
\chi_F&=N\cdot F\left(\rho, \frac{1}{N}\sum_j O_i^\dag O_j \rho O_j^\dag O_i\right)=N\left(\tr\sqrt{\frac{1}{N}\sum_j\sqrt{\rho} O_i^\dag O_j \rho O_j^\dag O_i\sqrt{\rho}}\right)^2\nonumber\\
&\leq\left(\sum_j\tr\sqrt{\sqrt{\rho} O_i^\dag O_j \rho O_j^\dag O_i\sqrt{\rho}}\right)^2=\left(\sum_j\sqrt{F}\left(\rho, O_i^\dag O_j \rho O_j^\dag O_i\right)\right)^2,
\end{align}
where the inequality holds from Eq. \eqref{eq: lemma2} and the fact that the matrix $\sqrt{\rho} O_i^\dag O_j \rho O_j^\dag O_i\sqrt{\rho}$ is positive semi-definite.

\section{\rom{3}. Derivation of Eq. \eqref{Eq: perturbed fidelity susceptibility} for $\Z_2$ order parameter}
Distinct from other charged local operators, the $\Z_2$ order parameter satisfies $O^\dag = O$. So for a $\Z_2$ order parameter, the corresponding fidelity susceptibility is slightly different, namely (again, we assume translation invariance)
\begin{align}
\chi_F=&\lim_{p\rightarrow0}\frac{1}{p}F\left((1-p)^N\rho + p(1-p)^{N-1}\sum_j O_j \rho O_j + p^2(1-p)^{N-2}\sum_{j,k}O_j O_k \rho O_k O_j, (1-p)^N O_i \rho O_i + p(1-p)^{N-1}\sum_j O_i O_j \rho O_j O_i\right)+\mathcal{O}(p)\nonumber\\
=&\lim_{p\rightarrow0}\frac{1}{p}F\left((1-p)^N\rho + p^2(1-p)^{N-2}\sum_{j,k}O_j O_k \rho O_k O_j, \ \ p(1-p)^{N-1}\sum_j O_i O_j \rho O_j O_i\right)\nonumber\\
&+\lim_{p\rightarrow0}\frac{1}{p}F\left(p(1-p)^{N-1}\sum_j O_j \rho O_j, \ \ (1-p)^N O_i \rho O_i\right)\nonumber\\
&+\lim_{p\rightarrow0}\frac{2}{p}\sqrt{F}\left((1-p)^N\rho + p^2(1-p)^{N-2}\sum_{j,k}O_j O_k \rho O_k O_j, p(1-p)^{N-1}\sum_j O_i O_j \rho O_j O_i\right)\sqrt{F}\left(p(1-p)^{N-1}\sum_j O_j \rho O_j, (1-p)^N O_i \rho O_i\right)\nonumber\\
=&\lim_{p\rightarrow0}(1-p)^{2N-1}F\left(\rho + \left(\frac{p}{1-p}\right)^2\sum_{j,k}O_j O_k \rho 
 O_k O_j, \sum_j O_i O_j \rho O_j O_i\right)+\lim_{p\rightarrow0}(1-p)^{2N-1}F\left(\sum_j O_j \rho O_j, O_i \rho O_i\right)\nonumber\\
 &+\lim_{p\rightarrow0}2(1-p)^{2N-1}\sqrt{F}\left(\rho + \left(\frac{p}{1-p}\right)^2\sum_{j,k}O_j O_k \rho 
 O_k O_j, \sum_j O_i O_j \rho O_j O_i\right)\sqrt{F}\left(\sum_j O_j \rho O_j, O_i \rho O_i\right).
\end{align}
In particular, we note that the first term of the above equation can be both upper and lower bounded simultaneously, namely
\begin{align}
F & \left(\rho + \left(\frac{p}{1-p}\right)^2\sum_{j,k}O_j O_k \rho 
 O_k O_j, \sum_j O_i O_j \rho O_j O_i\right)\nonumber\\
 & \geq F\left(\rho, \sum_j O_i O_j \rho O_j O_i\right) + \left(\frac{p}{1-p}\right)^2 F\left(\sum_{j,k}O_j O_k \rho O_k O_j, \sum_j O_i O_j \rho O_j O_i\right),
\end{align}
and 
\begin{align}
F & \left(\rho + \left(\frac{p}{1-p}\right)^2\sum_{j,k}O_j O_k \rho 
 O_k O_j, \sum_j O_i O_j \rho O_j O_i\right)\nonumber\\
& \leq \left[\sqrt{F}\left(\rho, \sum_j O_i O_j \rho O_j O_i\right) + \frac{p}{1-p}\sqrt{F}\left(\sum_{j,k}O_j O_k \rho O_k O_j, \sum_j O_i O_j \rho O_j O_i\right) \right]^2,
\end{align}
where we have applied Eq. \eqref{eq: lemma2}.
We notice that both upper and lower bounds converge to the fidelity $F\left(\rho, \sum_j O_i O_j \rho O_j O_i\right)$ in the limit $p\to 0$. Therefore, for $\Z_2$ order parameter, the fidelity susceptibility will have a prefactor 4, namely
\begin{align}
\chi_F^{\Z_2}=4N\cdot F\left(\rho, \frac{1}{N}\sum_j O_i O_j \rho O_j O_i\right).
\end{align}

\section{\rom{4}. Converging lower bounds for fidelity}
In this section, we provide the construction of a converging series of lower bounds, each consisting of polynomials of the density matrices, that approximates the fidelity between two density matrices. This improves the lower bound provided by Ref. \cite{miszczak2008sub}. 

We begin by expanding the fidelity as following:
\begin{equation}
F(\rho,\sigma)=\left[\tr\left(\sqrt{\sqrt{\rho}\sigma\sqrt{\rho}}\right)\right]^2=\left(\sum_i\sqrt{\lambda_i}\right)^2,
\end{equation}
where $\lambda_i$ is the $i$-th eigenvalue of the matrix $\sqrt{\rho}\sigma\sqrt{\rho}$. We note that, since $\sqrt{\rho}\sigma\sqrt{\rho}$ is similar to $\rho\sigma$ \cite{miszczak2008sub,baldwin2023efficiently}, $\{\lambda_i\}$ is also the set of eigenvalues of $\rho\sigma$. Expanding the square, we have
\begin{equation}
    \begin{aligned}
    F(\rho,\sigma)&=\sum_{i,j}\sqrt{\lambda_i}\sqrt{\lambda_j}=\sum_i\lambda_i+\sum_{i\neq j}\sqrt{\lambda_i}\sqrt{\lambda_j}
    \end{aligned}
\end{equation}
An increasing lower bound can be achieved as follows. Denoting $F\equiv F(\rho,\sigma)$ and $F_1\equiv \sum_i\lambda_i=\tr(\sqrt{\rho}\sigma\sqrt{\rho})=\tr(\rho\sigma)$, the first lower bound is simply
\begin{equation}
    F\geq F_1.
\end{equation}
To improve the lower bound, we calculate their difference:
\begin{equation}\label{eq: f-f1}
    \delta F_1\equiv (F-F_1)^2=\sum_{i\neq j,k\neq l}\sqrt{\lambda_i}\sqrt{\lambda_j}\sqrt{\lambda_k}\sqrt{\lambda_l}=2\sum_{i\neq j}\lambda_i\lambda_j+\sum_{\langle ij,kl\rangle}\sqrt{\lambda_i}\sqrt{\lambda_j}\sqrt{\lambda_k}\sqrt{\lambda_l},
\end{equation}
where $\langle ij,kl\rangle$ denotes the range of the indices to be $i\neq j,k\neq l/ \{i=k,j=l\text{ or }i=l,j=k\}$. To proceed, we define
\begin{equation}
F_2\equiv \sum_{i\neq j} \lambda_i\lambda_j =\left(\sum_i\lambda_i\right)^2-\sum_i\lambda_i^2 =\left[\tr(\rho\sigma)\right]^2-\tr\left[(\rho\sigma)^2\right].
\end{equation}
Then Eq. \eqref{eq: f-f1} gives a better lower bound of the fidelity, $\delta F_1\geq 2F_2$, which is obtained in Ref. \cite{miszczak2008sub}. Moving further, we have
\begin{equation}
    \delta F_2\equiv\left(\delta F_1-2F_2\right)^2=\sum_{\langle ij,kl\rangle, \langle mn,pq\rangle}\sqrt{\lambda_i}\sqrt{\lambda_j}\sqrt{\lambda_k}\sqrt{\lambda_l}\sqrt{\lambda_m}\sqrt{\lambda_n}\sqrt{\lambda_p}\sqrt{\lambda_q}=8\sum_{\langle ij,kl\rangle}\lambda_i\lambda_j\lambda_k\lambda_l+\cdots,
\end{equation}
where the ellipsis indicates the remaining terms that involve square roots. We similarly define
\begin{equation}
\begin{aligned}
    F_3&=\sum_{\langle ij,kl\rangle}\lambda_i\lambda_j\lambda_k\lambda_l=F_2^2-2\left[\left(\sum_{i}\lambda_i^2\right)^2-\sum_i\lambda_i^4\right]\\
    &=\left[\tr(\rho\sigma)\right]^4-2\left[\tr(\rho\sigma)\right]^2\tr\left[(\rho\sigma)^2\right]-\tr\left[(\rho\sigma)^2\right]^2+2\tr\left[(\rho\sigma)^4\right],
\end{aligned}
\end{equation}
so that $\delta F_2\geq 8F_3.$

It is clear from the above derivations that the inequality can be iterated to provide even closer lower bounds of fidelity. In general, for every $n>1$, we have the following recursion
\begin{equation}
    \delta F_n\equiv \left[\delta F_{n-1}-2^{\frac{n(n-1)}{2}}F_n\right]^2\geq 2^{\frac{n(n+1)}{2}}F_{n+1},
\end{equation}
where
\begin{align}
    F_{n+1}=F_{n}^2-2^{\frac{n(n-1)}{2}}\left[\left(\sum_{i}\lambda_i^{2^{n-1}}\right)^2-\sum_i\lambda_i^{2^{n}}\right].
\end{align}
Using these relations, the fidelity can be represented by the following exact form:
\begin{align}
    F=F_1+\sqrt{2F_2+\sqrt{8F_3+\sqrt{\cdots\sqrt{2^{\frac{n(n-1)}{2}}F_{n}+\sqrt{\cdots}}}}}.
\end{align}

In our case, the fidelity is between the system's density matrix $\rho$ and $\sigma=\frac{1}{N}\sum_j O_i^\dagger O_j\rho O_j^\dagger O_i$. Therefore, the fidelity proxy $F_{n}$ involves the R\'enyi-$2^{n}$ correlator 
\begin{align}
R^{(2n)}(i,j)=\tr\left[\left(\frac{1}{N}\sum_j\rho O_i^\dagger O_j\rho O_j^\dagger O_i\right)^{2^{n-1}}\right].
\end{align}

\end{document}